\documentclass[10pt]{iopart}

\usepackage{geometry}
\usepackage{array}
\usepackage{graphicx}
\usepackage{booktabs}
\usepackage{gensymb}
\usepackage{hyperref}
\hypersetup{
    colorlinks=true,
    linkcolor=blue,
    filecolor=magenta,      
    urlcolor=blue,
    citecolor=blue,
    pdftitle={Overleaf Example},
    pdfpagemode=FullScreen,
    }
\expandafter\let\csname equation*\endcsname\relax
\expandafter\let\csname endequation*\endcsname\relax
\usepackage{amsmath,amssymb}
\usepackage[
backend=biber,
style=phys,
maxnames=1,
eprint=true,
]{biblatex}
\DeclareSourcemap{
  \maps[datatype=bibtex]{
    \map{% insert missing spaces between initials ("A.V." -> "A. V.") so all initials are printed
      \step[fieldsource=author, match=\regexp{(\p{Lu}\.)(\p{Lu})}, replace={$1 $2}]
      \step[fieldsource=author, match=\regexp{(\p{Lu}\.)(\p{Lu})}, replace={$1 $2}]
    }
    \map{
      \step[fieldset=language, null]
    }
    \map{% Zotero writes presentation presenters as an internal note ("Presenters: _:n12345")
      \step[fieldsource=note, match=\regexp{^Presenters:}, final]
      \step[fieldset=note, null]
    }
    \map{% remove Chinese characters from author names (pdfLaTeX cannot typeset them)
      \step[fieldsource=author, match=\regexp{\s*\p{Han}+}, replace={}]
    }
    \map{
      \step[fieldsource=doi, matchi=\regexp{^10\.48550/arxiv\.(.+)$}, final]
      \step[fieldset=eprint, fieldvalue={$1}]
      \step[fieldset=eprinttype, fieldvalue={arxiv}]
      \step[fieldset=doi, null]
      \step[fieldset=url, null]
      \step[fieldset=note, null]
    }
  }
}

\AtEveryBibitem{%
  \ifboolexpr{ test {\iffieldundef{journaltitle}}
               and test {\iffieldequalstr{eprinttype}{arxiv}} }
    {\clearfield{title}\clearfield{eprintclass}\clearfield{year}\clearfield{month}}
    {}}
\renewbibmacro*{note+pages}{%
  \printfield{note}%
  \setunit{\bibpagespunct}%
  \printfield{pages}%
  \iffieldundef{pages}
    {\iffieldundef{pubstate}{\printfield{doi}}{}%
     \clearfield{doi}}
    {\iftoggle{bbx:doi}{}{\clearfield{doi}}}}

\DeclareFieldFormat[thesis]{type}{\ifbibstring{#1}{\bibstring{#1}}{\bibstring{phdthesis}}}
\DeclareFieldFormat[thesis]{doi}{doi\addcolon\href{https://doi.org/#1}{\nolinkurl{#1}}}
\AtEveryBibitem{%
  \ifentrytype{thesis}
    {\clearfield{title}\clearlist{location}\clearfield{month}}
    {}}
\renewbibmacro*{institution+location+date}{%
  \ifentrytype{thesis}
    {\setunit{\addcomma\space}%
     \printlist{institution}%
     \setunit{\addcomma\space}%
     \usebibmacro{date}%
     \setunit{\addcomma\space}%
     \printfield{doi}%
     \clearfield{doi}}
    {\setunit{\addspace}%
     \printtext[parens]{%
       \printlist{institution}%
       \newunit
       \printlist{location}%
       \newunit
       \usebibmacro{date}}}}

\AtEveryBibitem{%
  \ifkeyword{conference}
    {\clearfield{title}\clearfield{month}\clearfield{url}}
    {}}
\usepackage{blindtext}
\begin{document}

\title[High-power X-Point Target divertor experiments on the TCV tokamak]{High-power X-Point Target divertor experiments on the TCV tokamak}

\author{K.~Lee$^1$, M.~Carpita$^1$, C.~Theiler$^1$, M.~Zurita$^1$, O.~Février$^1$,  A.~Mele$^1$, R.I.~Morgan$^1$, F.~Pastore$^{1,2}$, H.~Reimerdes$^1$, K.~Verhaegh$^3$, M.~Winkel$^{3,4}$, D.~Brida$^2$, B.Y.K.~Brown$^1$, M.J.H. Cornelissen$^{1,3,4}$, R.~Ducker$^1$, G.~Durr-Legoupil-Nicoud$^1$, D.~Hamm$^1$, M.~Pedrini$^1$, A.~Perek$^1$, O.~Sauter$^1$, P.~Sintre$^1$, E.~Tonello$^1$, Y.~Wang$^1$, the TCV Team$^{\dag}$ and the EUROfusion Tokamak Exploitation Team$^{\ddag}$}

\address{$^1$ \'Ecole Polytechnique F\'ed\'erale de Lausanne (EPFL), Swiss Plasma Center (SPC), CH-1015 Lausanne, Switzerland}
\address{$^2$ Max Planck Institute for Plasma Physics, Garching, Germany}
\address{$^3$ Department of Applied Physics and Science Education, Science and Technology of Nuclear Fusion, Eindhoven University of Technology, Eindhoven, Netherlands}
\address{$^4$ DIFFER - Dutch Institute for Fundamental Energy Research, Eindhoven, Netherlands}
\address{$\dag$ See author list of C. Theiler, et al., Nucl. Fusion 66 (2026) 116007.}
\address{$\ddag$ See author list of N. Vianello, et al., Nucl. Fusion 66 (2026) 116010.}

\ead{kenneth.lee@epfl.ch}
\vspace{10pt}

\begin{abstract}
Stationary power exhaust in the X-Point Target (XPT) divertor configuration is investigated in high-power ($2.5\,\text{MW}$) TCV L-mode discharges, in which the Lengyel detachment scaling metric $P_{\text{SOL}}B_{0}/(R_{0}n_{e,\text{u}}^{2})$ lies in the range expected for reactors (SPARC, ITER, ARC). Compared with a conventional single-null (SN) divertor, the XPT improves power exhaust performance, reducing the peak outer target heat flux by 40--60\% and enabling divertor detachment with nitrogen seeding under conditions in which the SN cannot be detached. Detachment evolves asymmetrically across the multiple XPT outer strike points. The sensitivities of the peak target heat fluxes and impurity emission front positions to the X-point separation ($\text{d}R_{\text{u},X2}$) are characterized and found to decrease at higher nitrogen seeding levels. Field-reversal experiments reveal the critical role of $E\times B$ drifts inherent to the magnetic topology of the secondary X-point, which drive an unconventional redistribution of density and impurity radiation in the XPT divertor and thereby cause the asymmetries between the XPT outer strike points, largely confirming the expectations from recent SOLPS-ITER modelling. A framework for understanding the power exhaust physics of the XPT is presented, detailing the roles of SOL splitting, connection length peaking, $E\times B$ drifts, and vessel geometry. Together, these results establish an experimental physics basis for the XPT divertor as a potential power exhaust solution for tokamak reactors.

\end{abstract}

%
% Uncomment for keywords
\vspace{2pc}
\noindent{\it Keywords}: power exhaust, alternative divertor configurations, X-Point Target divertor, detachment, drifts, TCV
%
% Uncomment for Submitted to journal title message
%\submitto{\JPA}
%
% Uncomment if a separate title page is required
\maketitle
% 
% For two-column output uncomment the next line and choose [10pt] rather than [12pt] in the \documentclass declaration
\ioptwocol

\section{Introduction}
% Introduce the power exhaust challenge and the need for alternatives
Next-step, gigawatt-scale tokamak fusion power plants will require divertor power exhaust solutions that integrate core performance with plasma detachment \cite{leonard_plasma_2018,krasheninnikov_physics_2017} under divertor heat fluxes higher than those of ITER \cite{wenninger_demo_2014,loarte_power_2017,zohm_eu_2021,reimerdes_assessment_2020,eich_power_2026}. Alternative divertor configurations \cite{soukhanovskii_review_2017,theiler_results_2017,verhaegh_divertor_2025} have been conceived to tackle this extreme power exhaust challenge and to mitigate the extrapolation risks of the ITER-like conventional single-null (SN) divertor \cite{pitts_physics_2019}. Through advanced magnetic shaping of the divertor, these alternative configurations are expected to reduce the separatrix density and impurity seeding \cite{kallenbach_impurity_2013} required to attain detachment, and to widen the parameter window for detached divertor operation \cite{lipschultz_sensitivity_2016,cowley_optimizing_2022}.

The X-Point Target (XPT) divertor \cite{labombard_x-point_2013,labombard_adx_2015} is a promising alternative divertor configuration characterized by a secondary X-point along the divertor leg. First realized experimentally in TCV Ohmic L-mode plasmas \cite{theiler_results_2017}, the XPT has been found to greatly reduce the upstream separatrix density threshold for detachment access \cite{lee_x_2025}. The impurity emission front clamps at the secondary X-point, yielding a wide detachment operational window \cite{lee_x_2025} that is resilient against fast perturbations in fuelling, seeding and heating \cite{winkel_detachment_2026}. In H-mode discharges with type-I edge-localized modes (ELMs), a reduction of target heat fluxes by the XPT has been observed in the inter-ELM phase \cite{raj_improved_2022} and, more recently, during the ELM crash \cite{zurita_mitigation_2025,zurita_elm_2026}. The large poloidal flux expansion at the secondary X-point is thought to be responsible for these exhaust benefits \cite{lee_x_2025,lipschultz_sensitivity_2016}, which are related to the phenomenon of X-point radiation \cite{bernert_x-point_2021,stroth_model_2022,bernert_x-point_2025}. 

The XPT is currently being considered as an alternative divertor candidate for next-step devices (SPARC \cite{kuang_divertor_2020}, ARC \cite{eich_power_2026}, EHL-2 \cite{wang_divertor_2025}). In particular, the ARC fusion power plant is designed with up-down symmetric (double-null) XPT divertors in long, tightly baffled outer legs. The engineering concepts required to implement the XPT in a practical reactor are being developed \cite{kuang_conceptual_2018}, including the creation of additional divertor volume within the breeding blanket to accommodate long divertor legs, tight baffling for neutral compression and power dissipation via plasma-neutral interactions \cite{umansky_study_2020,reimerdes_implementation_2026}, and demountable high-temperature superconducting (HTS) magnets \cite{sorbom_arc_2015} enabling the placement of divertor shaping coils within the toroidal field coil volume in reactor environments. Equilibrium design and magnet control schemes for the XPT are also being studied \cite{wai_equilibrium_2021,bardsley_decoupled_2024,ding_advantage-level_2026}. Such a strongly shaped, high-performance divertor may be particularly needed in compact, high-field devices such as ARC \cite{eich_power_2026}, which are projected to produce extreme unmitigated divertor heat fluxes ($q_{\parallel}\sim P_{\text{SOL}}B_{0}/R_{0}$ \cite{labombard_adx_2015,eich_inter-elm_2011}).

Reliable extrapolation of the XPT to a reactor requires both physics understanding and predictive capability. Alongside experimental proof-of-principle and validation (TCV \cite{lee_x_2025}, MAST-U \cite{lonigro_initial_2026}, EXL-50U \cite{ding_advantage-level_2026}), high-fidelity modelling efforts are advancing rapidly \cite{umansky_assessment_2017,wigram_performance_2019,cowley_novel_2023,wigram_modelling_2025,muscente_investigating_2025,garcia_herreros_study_2026,wu_solps-iter_psi_2026}. Recent first-of-a-kind SOLPS-ITER simulations of the TCV XPT divertor including drifts and currents \cite{carpita_physics_2026,carpita_investigation_2026,carpita_inprep} indicate that divertor power dissipation is closely linked to $E \times B$ drifts intrinsic to the secondary X-point.

This paper investigates XPT power exhaust in a recently developed high-power TCV scenario \cite{lee_high-power_2026} in order to explore experimentally the SOL regime in which the power exhaust challenge is maximized. The highest available ECRH power ($2.5\,\text{MW}$, fully absorbed in a $\sim 1 \,\text{m}^{3}$ plasma) is applied at low upstream separatrix densities ($n_{e,\text{u}}\approx 1\times10^{19} \,\text{m}^{-3}$). Assuming a simple scaling of the impurity fraction required for detachment, quantified by the Lengyel metric $P_{\text{SOL}}B_{0}/(R_{0}n_{e,\text{u}}^{2})$ \cite{body_detachment_2024}, the detachment access challenge approaches that expected for reactors (SPARC, ITER, ARC). With nitrogen seeding, detachment is obtained in the XPT, whereas the SN cannot be detached. The sensitivities of the peak target heat fluxes to the X-point separation are characterized and found to decrease with nitrogen seeding. Transport due to $E \times B$ drifts inherent to the magnetic topology of the secondary X-point drives substantial changes in the density and impurity radiation distributions and affects the power dissipation in the XPT, confirming the expectations from SOLPS-ITER modelling \cite{carpita_physics_2026,carpita_investigation_2026,carpita_inprep}. To interpret these results, a framework is introduced that describes how the XPT geometry governs power exhaust and clarifies the role of the X-point separation, pointing to pathways for its geometric optimization.

This paper is organized as follows. Section \ref{section:xpt-shape} presents a framework for interpreting the SOL and power exhaust physics arising from the geometry of the XPT, detailing the roles of SOL splitting, connection length peaking, $E\times B$ drifts around the secondary X-point, and vessel geometry. Section \ref{section:setup} summarizes the experimental setup of the high-power XPT discharges with shape feedback control. Section \ref{section:attached} describes the properties of an attached XPT divertor, including the observation of density redistribution, the heat flux profiles across the multiple XPT strike points, and their sensitivity to X-point separation. Section \ref{section:detachment} presents observations during nitrogen seeding, including the demonstration of detachment in the XPT and the occurrence of asymmetric detachment evolution. Section \ref{section:drifts} describes the effect of field reversal on XPT power exhaust and identifies $E\times B$ drifts as the origin of the divertor density redistribution. Section \ref{section:discussion} discusses the implications for reactors, and section \ref{section:conclusion} summarizes the conclusions.

\section{SOL and power exhaust in the X-Point Target divertor geometry}\label{section:xpt-shape}

\begin{figure*}
    \centering
    \includegraphics[width=1\linewidth]{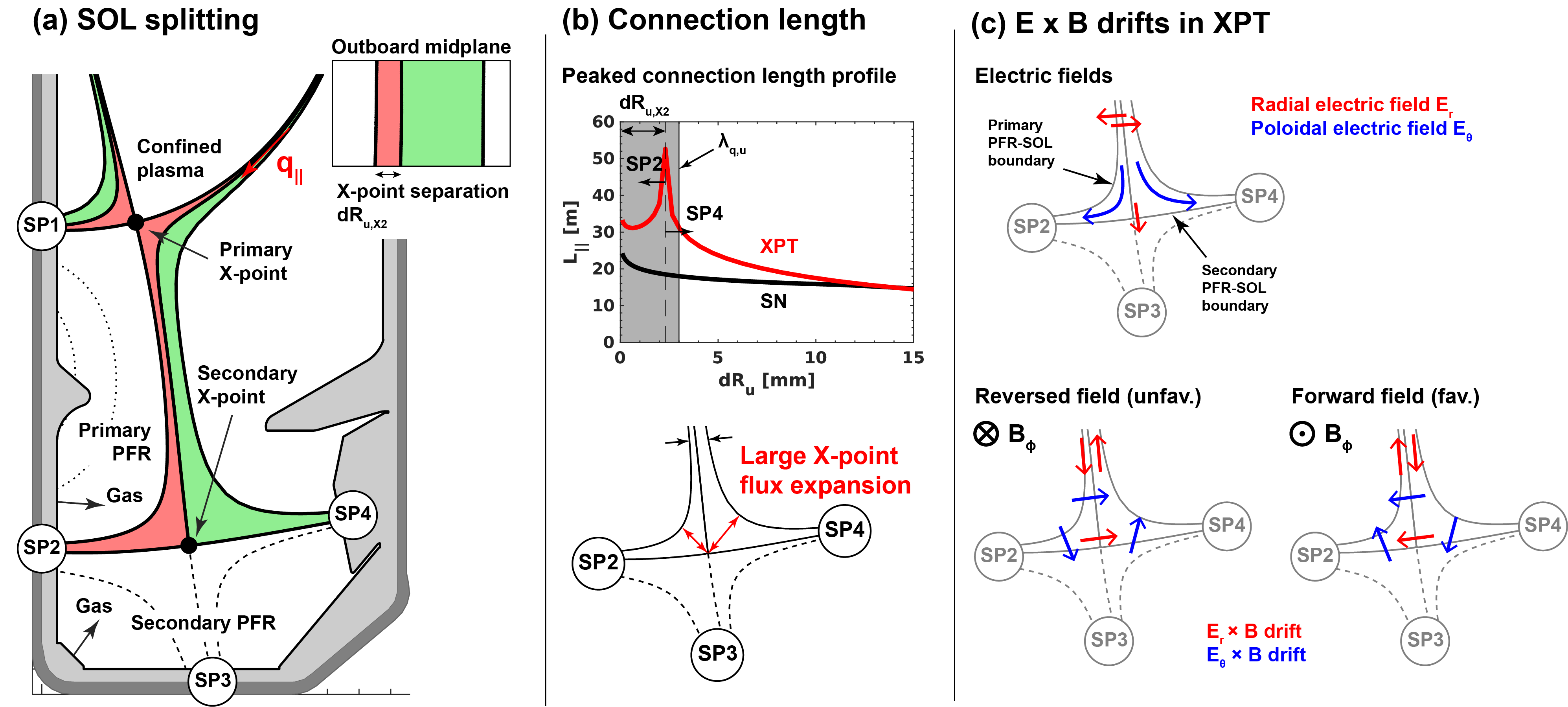}
    \caption{Sketches of the physics mechanisms underlying the power exhaust of an X-Point Target divertor, namely (a) heat flux sharing by SOL splitting, (b) connection length peaking at the secondary X-point, and (c) $E \times B$ drifts around the secondary X-point in the XPT. SOL quantities are indexed by ``u'' for upstream (outboard midplane) and ``t'' for target, with no index for a general point along a SOL field line. The radial flux coordinate $\text{d}R_{\text{u}}=R_{\text{u}}-R_{\text{u,sep}}$, i.e.\ the distance between a flux surface and the primary separatrix evaluated at the outboard midplane, is used to label each poloidal flux surface in the edge and SOL.}
    \label{fig:xpt_physics}
\end{figure*}

The magnetic geometry concepts of the XPT are reviewed here to provide a framework for interpreting the expectations and results in terms of SOL and power exhaust physics. A summary is shown in figure \ref{fig:xpt_physics}, using the TCV tokamak as an example. An XPT divertor is characterized by a secondary X-point embedded along the outer divertor leg, giving it the same magnetic topology as a snowflake divertor \cite{ryutov_snowflake_2015,gorno_power_2023,reimerdes_access_2024}. The difference is that the snowflake divertor places the two X-points close to each other near the main plasma (a quasi-second-order null), thereby creating a large region of very weak poloidal magnetic field $B_{\theta}$. In the XPT, the secondary X-point is placed far from the primary X-point to separate the divertor radiation from the core \cite{lee_x_2025}, and the SOL has already broadened along the divertor leg by the time the plasma reaches the secondary X-point.

The secondary X-point in the divertor leg splits the SOL, forming multiple strike points (SPs), labelled SP1--4 following TCV's counter-clockwise naming convention for divertors with multiple X-points (figure \ref{fig:xpt_physics}(a)). SP1 refers to the inner strike point. The outer strike points SP2 and SP4 are magnetically connected to the upstream SOL and receive the majority of the exhaust plasma fluxes. SP3 is upstream-disconnected and located in the secondary private flux region (PFR) below the secondary X-point, and it generally receives a small amount of plasma flux via cross-field transport across the secondary separatrix. The secondary PFR also offers opportunities to inject gas (fuelling or impurity seeding) directly into the high heat flux portion of the SOL through the secondary PFR-SOL boundary \cite{carpita_physics_2026,carpita_investigation_2026,carpita_inprep}.

A key shaping parameter is the flux coordinate of the secondary X-point, $\text{d}R_{\text{u},X2}$, shown in figure \ref{fig:xpt_physics}(a) and referred to as the ``X-point separation''\footnote{A similar quantity, $\text{d}R_{\text{sep}}$, is used to quantify the up-down magnetic balance in double-null plasmas. Note that $\text{d}R_{\text{sep}}$ and $\text{d}R_{\text{u},X2}$ measure the X-point separation in flux space, not the distance between the X-points in the poloidal plane.}. The optimal $\text{d}R_{\text{u},X2}$ is defined as the value that minimizes the peak target parallel heat flux $q_{\parallel,\text{t}}$ across both active SPs (here, SP2 and SP4). It is typically assumed that $\text{d}R_{\text{u},X2}$ should be set to a fraction of the SOL heat flux decay width $\lambda_{q,\text{u}}$ to optimize XPT performance \cite{labombard_adx_2015,wigram_performance_2019,eich_power_2026}, i.e.,
\begin{equation}\label{eq:dRx2_crit}
    \text{d}R_{\text{u},X2} \lesssim \lambda_{q,\text{u}}.
\end{equation}

The experimental results in this paper show that the role of the X-point separation $\text{d}R_{\text{u},X2}$ is highly nuanced. Flux tubes connected to each strike point can undergo very different processes, so that the final heat flux profiles vary with vessel geometry (section \ref{section:attached}), impurity seeding level (section \ref{section:detachment}), and field direction (section \ref{section:drifts}). All of these aspects affect (1) the optimal $\text{d}R_{\text{u},X2}$ and (2) the impact of deviating from this optimal value (i.e.\ the sensitivity of power exhaust to $\text{d}R_{\text{u},X2}$). In the remainder of this section, the power exhaust mechanisms governed by $\text{d}R_{\text{u},X2}$ are discussed in detail, allowing these conceptual subtleties to be addressed explicitly.

\subsection{SOL splitting: heat flux sharing and diffusive spreading}\label{subsection:SOL_splitting}
The first role of the X-point separation is to set the degree of SOL splitting (figure \ref{fig:xpt_physics}(a)). SOL flux tubes with $\text{d}R_{\text{u}}<\text{d}R_{\text{u},X2}$ are connected to SP2, and those with $\text{d}R_{\text{u}}>\text{d}R_{\text{u},X2}$ to SP4. Since the majority of the divertor heat flux is assumed to be carried within the first $\lambda_{q,\text{u}}$, meaningful SOL splitting, and hence heat flux sharing between multiple strike points, is expected when $\text{d}R_{\text{u},X2} \lesssim \lambda_{q,\text{u}}$. However, for purely parallel transport in the absence of volumetric power losses, SOL splitting by itself does not reduce the peak $q_{\parallel,\text{t}}$ for any $\text{d}R_{\text{u},X2}$.

A reduction in peak $q_{\parallel,\text{t}}$ is achieved through diffusive spreading of the heat flux in the divertor \cite{eich_inter-elm_2011,maurizio_effect_2019}, with the optimum then determined by magnetically balancing the peak heat flux between the two strike points. Diffusive spreading into the secondary PFR can also activate the disconnected SP3. A detailed discussion of these assumptions (mainly in the context of a snowflake divertor), together with simple modelling, can be found in \cite{labit_experimental_2017,lunt_numerical_2016,lunt_proposal_2017}.

\subsection{Connection length peaking: SOL cooling, power losses and detachment}\label{subsection:cl_peaking}
The second role of the X-point separation is to determine which portion of the SOL experiences a significant increase in parallel connection length $L_{\parallel}$ (figure \ref{fig:xpt_physics}(b)). In the XPT configuration, the $L_{\parallel}$ profile diverges ($L_{\parallel} \rightarrow \infty$) at $\text{d}R_{\text{u}}=\text{d}R_{\text{u},X2}$, since the magnetic field is purely toroidal at an X-point. The locally peaked profile increases the total connection length of SOL flux tubes near the secondary X-point relative to the SN configuration. These flux tubes also experience large poloidal flux expansion (the expansion of the perpendicular flux surface spacing relative to upstream) due to the locally weak poloidal field.

Analytical and numerical models of the divertor SOL show that increasing the connection length $L_{\parallel}$ can promote target cooling (two-point model \cite{stangeby_plasma_2000}), lower the detachment threshold (Lengyel model \cite{lengyel_analysis_1981,goldston_new_2017}, 1D Braginskii \cite{gorno_x-point_2024,fevrier_splend1d_2024}) and reduce the sensitivity of the detachment front location to control parameters (DLS model \cite{lipschultz_sensitivity_2016,cowley_optimizing_2022}). Given these expected power exhaust benefits of long connection lengths, it has been argued that the ``ideal'' XPT should place the power-carrying flux tubes at $L_{\parallel}\rightarrow \infty$ \cite{labombard_adx_2015,eich_power_2026,lee_x_2025} by positioning the secondary X-point according to $\text{d}R_{\text{u},X2} \lesssim \lambda_{q,\text{u}}$.

Although the conclusion is similar to that of the SOL splitting argument in section \ref{subsection:SOL_splitting}, the argument here is based on optimizing the SOL connection length profile for detachment, and therefore follows a different line of reasoning. A similar conclusion was drawn from a $\text{d}R_{\text{u},X2}$ scan in UEDGE modelling of the ARC XPT divertor without drifts \cite{wigram_performance_2019}. The efficiency of power dissipation, however, depends strongly on the density in the SOL flux tubes. Cases have been found in which $E\times B$ drifts (section \ref{subsection:ExB_modified}) displace particles from the $L_{\parallel}$-peaked flux tubes, which can undermine this mechanism.

\subsection{$E \times B$ drifts: convective transport around the secondary X-point}\label{subsection:ExB_modified}
The third role of the X-point separation is to influence the behaviour of $E \times B$ drifts \cite{chankin_classical_1997}, as first predicted by SOLPS-ITER modelling \cite{carpita_physics_2026,carpita_investigation_2026,carpita_inprep} and summarized here. The XPT can exhibit $E \times B$ drift patterns that differ from those of the conventional SN, especially around the secondary X-point as a consequence of its magnetic topology, thereby introducing particle convection and modifying the 2D profile of radiative losses. More precisely, the X-point separation determines which flux tubes are directly exposed to the secondary PFR, where the potential profile across the secondary PFR-SOL boundary influences the drift strength and convection channels.

Figure \ref{fig:xpt_physics}(c) illustrates the two components of the $E \times B$ drift \cite{stangeby_plasma_2000}, namely the $E_{r} \times B$ drift, which points poloidally along the SOL flux surfaces, and the $E_{\theta} \times B$ drift, which points radially across them. The $E_{r} \times B$ drift is typically larger in magnitude than the $E_{\theta} \times B$ drift because of the steep potential gradients at the PFR-SOL boundary \cite{tsui_relevance_2022,de_oliveira_new_2022}. Above the secondary X-point, the $E \times B$ drift patterns are those normally expected in an SN, but distinct features emerge around the secondary X-point. The first is an $E_{r} \times B$ drift that drives \emph{radial} flow across flux surfaces right above the secondary X-point, providing an additional cross-field transport mechanism between the SP2 and SP4 manifolds. This component arises because the bulk SOL flux tubes open onto a secondary PFR-SOL boundary, which has a different potential structure and can develop strong, localized radial electric fields pointing into the secondary PFR, with an amplitude that also varies with field direction in the presence of Pfirsch--Schl\"uter currents (X-point potential well \cite{wensing_x-point_2020,carpita_physics_2026,carpita_investigation_2026,carpita_inprep}). The large flux expansion at the X-point further enhances this effect, (1) through large localized power losses, which increase the plasma cooling \cite{lee_x_2025} and hence the potential variations, and (2) by increasing the relative importance of cross-field drifts through the small field-line angles. The second feature is the $E_{\theta} \times B$ drift, which can convect particles into or out of the upstream-disconnected secondary PFR. These XPT-specific drifts act predominantly on SOL flux tubes near the secondary PFR-SOL boundary, whose position is determined by $\text{d}R_{\text{u},X2}$.

Radial drift convection mainly affects particle transport rather than power transport because parallel heat conduction remains efficient \cite{carpita_physics_2026,carpita_investigation_2026,carpita_inprep}. This leads to larger momentum losses relative to power losses in SOL flux tubes, which reduces the plasma density and increases the plasma temperature ($n_{e,\text{t}}\propto (1-f_{\text{mom}})^{3}$ and $T_{e,\text{t}}\propto (1-f_{\text{mom}})^{-2}$ in the two-point model \cite{stangeby_plasma_2000}, with $f_{\text{mom}}$ the momentum loss factor). The plasma density is lowered and the temperature raised in one of the SP2 and SP4 manifolds, while the density is raised and the temperature lowered in the other. Impurity ions also experience $E \times B$ drifts and, together with the bulk plasma, modify the poloidal distribution of impurity radiation. The power dissipation efficiency in individual SOL flux tubes changes accordingly, since the radiated power is proportional to $n_{e}^{2}$ at fixed impurity concentration \cite{lengyel_analysis_1981}. As a result, drift-driven asymmetry between SP2 and SP4 emerges and depends on the toroidal field direction.

The impact of drifts on power exhaust performance \cite{carpita_physics_2026,carpita_investigation_2026,carpita_inprep} is central to the interpretation of the results in this work. Experimental evidence of convective transport originating from $E\times B$ drifts is presented in section \ref{section:drifts}.

\subsection{Vacuum vessel geometry: strike-point positioning and plasma-wetted area}\label{subsection:Rt_effect}
Beyond the mechanisms discussed above, the X-point separation affects power exhaust indirectly via the radial position of the strike points. The parallel heat flux transported from upstream to the target is compressed or diluted by variations in the cross-sectional area of the flux tubes, according to
\begin{equation}\label{eq:Rt_effect}
    q_{\parallel,\text{t}} = q_{\parallel,\text{u}} (1-f_{\text{pow}}) \frac{B_{\text{t}}}{B_{\text{u}}} \approx q_{\parallel,\text{u}} (1-f_{\text{pow}}) \frac{R_{\text{u}}}{R_{\text{t}}}.
\end{equation}
Here, $f_{\text{pow}}$ is the power loss fraction along the flux tube due to volumetric dissipation or cross-field spreading, and the factor $R_{\text{u}}/R_{\text{t}}$ accounts for flux tube compression or dilation. This factor is the inverse of the target total flux expansion, $f_{R,\text{t}}=B_{\text{u}}/B_{\text{t}}\approx R_{\text{t}}/R_{\text{u}}$\footnote{In some works, the total flux expansion is defined between the primary X-point (instead of upstream) and the target.}. This effect is particularly important for TCV, whose vacuum vessel geometry typically places the XPT SP2 on the inner wall, where geometric compression alone increases $q_{\parallel}$ from upstream to target, and must therefore be factored into the choice of the optimal $\text{d}R_{\text{u},X2}$ (illustrated in section \ref{subsection:attached_dRx2}). This is less of an issue in most XPT devices (e.g.\ MAST-U \cite{lonigro_initial_2026}, SPARC \cite{kuang_divertor_2020}, ARC \cite{eich_power_2026}) where the outer divertor leg is radially extended and SP2 lies at a larger radius.

\section{Experimental setup}\label{section:setup}

\begin{figure*}
    \centering
    \includegraphics[width=1\linewidth]{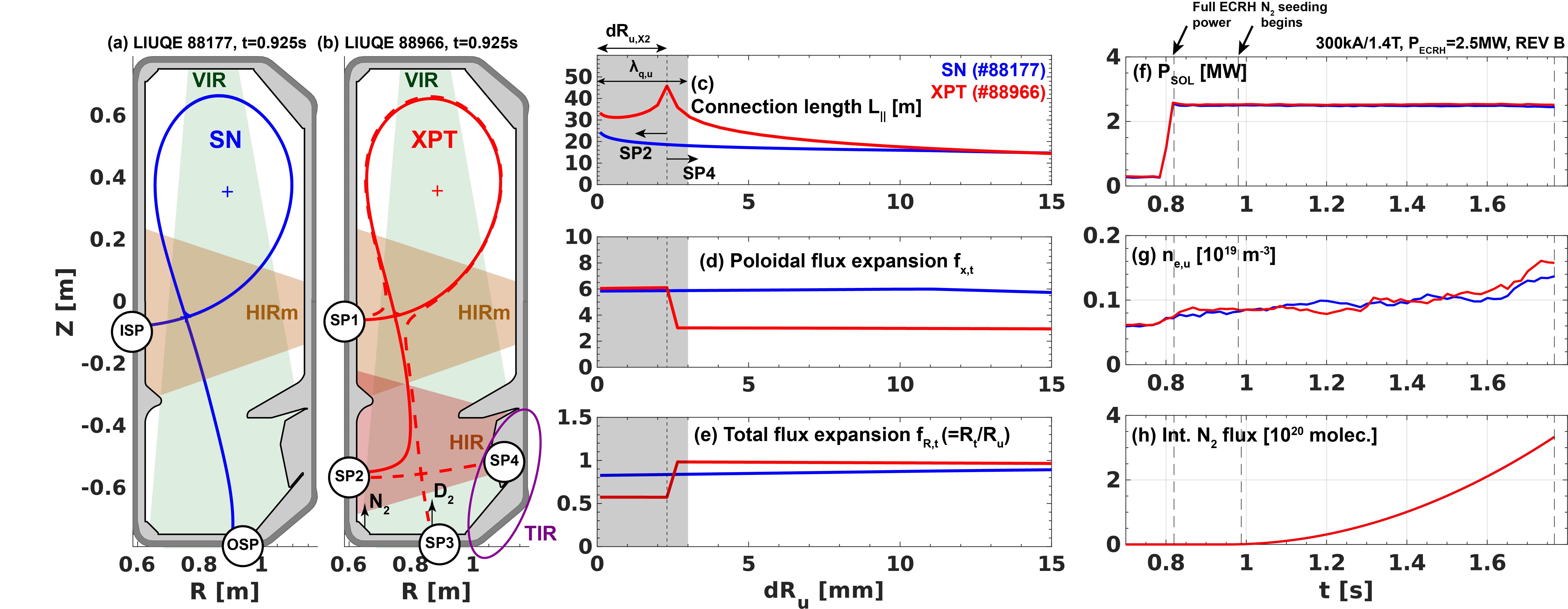}
    \caption{TCV poloidal cross-sections showing the (a) SN (\#88177) and (b) XPT (\#88966) configurations studied in this work. Highlighted are all strike point positions (OSP and ISP denote the outer and inner strike points in the SN) and the fields of view of the vertical (VIR), horizontal (HIR), horizontal midplane (HIRm) and tangential (TIR) infrared cameras. The gas injection locations are also shown. Radial profiles of (c) connection length, (d) target poloidal flux expansion and (e) target total flux expansion, with all geometric quantities evaluated between the outboard midplane and the outer target. Time traces of (f) the power entering the SOL, (g) the upstream separatrix density and (h) the time-integrated nitrogen flux.}
    \label{fig:eqm}
\end{figure*}

\subsection{High-power L-mode scenario}
The XPT divertor experiments were performed on the TCV tokamak \cite{hofmann_creation_1994,theiler_progress_2026} ($R_{0}=0.88\,\text{m}$, $B_{0}=1.4\,\text{T}$, full carbon first wall), using a recently developed high-power, low-density L-mode scenario (described in detail in \cite{lee_high-power_2026}) at $I_{\text{P}}=300\,\text{kA}$ and $q_{95}\approx 2.5$. The heat flux decay width is estimated to be $\lambda_{q,\text{u}}\approx 3\,\text{mm}$ \cite{lee_high-power_2026}. An overview of the experimental setup is shown in figure \ref{fig:eqm}. Figure \ref{fig:eqm}(f)--(h) shows time traces of key power exhaust parameters for an SN and an XPT discharge with the highest available ECRH power ($2.5\,\text{MW}$ in second-harmonic X-mode at $\sim100\%$ absorption) and divertor power dissipation induced by a nitrogen seeding ramp. The power entering the SOL, $P_{\text{SOL}}\approx 2.5\,\text{MW}$ (figure \ref{fig:eqm}(f)), is estimated as the difference between the heating power (absorbed ECRH plus Ohmic) and the core radiated power. The density is feedback controlled by adjusting the deuterium fuelling rate using either the line-averaged density measured by the far-infrared interferometer (FIR) or the model-based real-time density profile observer RAPDENS-EKF \cite{pastore_applications_2026}. Nitrogen is injected in feed-forward from the divertor up to a time-integrated flux of $3.3\times10^{20} \,\text{molec.}$ (figure \ref{fig:eqm}(h)). Over this period, the line-averaged density within the LCFS varies over $\langle n_{e}\rangle \approx 1.3\text{--}2.3 \times 10^{19}\, \text{m}^{-3}$ (corresponding to Greenwald fractions $f_{\text{G}}\approx 0.06\text{--}0.1$) and the upstream separatrix density over $n_{e,\text{u}}\approx0.8\text{--}1.5\times10^{19}\,\text{m}^{-3}$ (figure \ref{fig:eqm}(g)). Unless stated otherwise, the experiments are carried out with the toroidal field in the reversed (H-mode unfavourable) direction.

%field direction, with the ion $\nabla B$ drift pointing away from the primary X-point, labelled REV B) is the main scenario discussed in this paper (sections \ref{section:attached} and \ref{section:detachment}). Operation in forward field (favourable field direction, with the ion $\nabla B$ drift pointing towards the primary X-point, labelled FWD B) is also achieved without an H-mode transition, since the density lies on the ``low-density branch'' of the L-H transition threshold \cite{labit_lh_2025}; these results are discussed in section \ref{section:drifts}.

The divertor impurity fraction needed for detachment can be quantified by the Lengyel metric $P_{\text{SOL}}B_{0}/(R_{0}n_{e,\text{u}}^{2})$ \cite{body_detachment_2024}. The present high-power L-mode scenario \cite{lee_high-power_2026} reaches a Lengyel metric of $\sim200\text{--}300$\footnote{The difference in $\lambda_{q,\text{u}}$ between L- and H-mode has been accounted for by including a factor of 2. See details in \cite{lee_high-power_2026}.}, a factor of $10$ larger than in the well-studied Ohmic L-mode \cite{theiler_results_2017,fevrier_nitrogen-seeded_2020} and NBI H-mode \cite{raj_improved_2022} scenarios and comparable to those of reactors (SPARC, ITER and ARC). Combined with the divertor shaping flexibility of TCV, these conditions enable experimental study of XPT power exhaust under relatively high divertor heat fluxes ($q_{\parallel}\sim100\,\text{MW m}^{-2}$) and in SOL regimes where detachment is difficult to achieve.

\subsection{Divertor geometry and magnetic control}
The SN and XPT configurations considered in this work are shown in detail in figure \ref{fig:eqm}(a) and (b). The gas baffles \cite{reimerdes_initial_2021,fevrier_divertor_2021,reimerdes_fec_2023} are in the SI-LO (Short-Inner, Long-Outer) configuration. Equilibria are reconstructed in real time using the LIUQE code \cite{moret_tokamak_2015}. The plasmas are positioned near the top of the vessel to allow a long, variable divertor leg while minimizing changes to the core shape. The X-point separation shown in figure \ref{fig:eqm}(b) is $\text{d}R_{\text{u},X2}\approx2.2\,\text{mm}$, which increases the connection length between the outboard midplane and the target by approximately a factor of 2 relative to the SN over most of the first $\lambda_{q,\text{u}}$ (figure \ref{fig:eqm}(c)). Although the poloidal flux expansion is large at the secondary X-point, this does not carry through to the target (figure \ref{fig:eqm}(d)). The total flux expansion $f_{R,\text{t}}$ (figure \ref{fig:eqm}(e)) modifies the parallel heat flux between upstream and target according to equation (\ref{eq:Rt_effect}).

\begin{figure}
    \centering
    \includegraphics[width=0.8\linewidth]{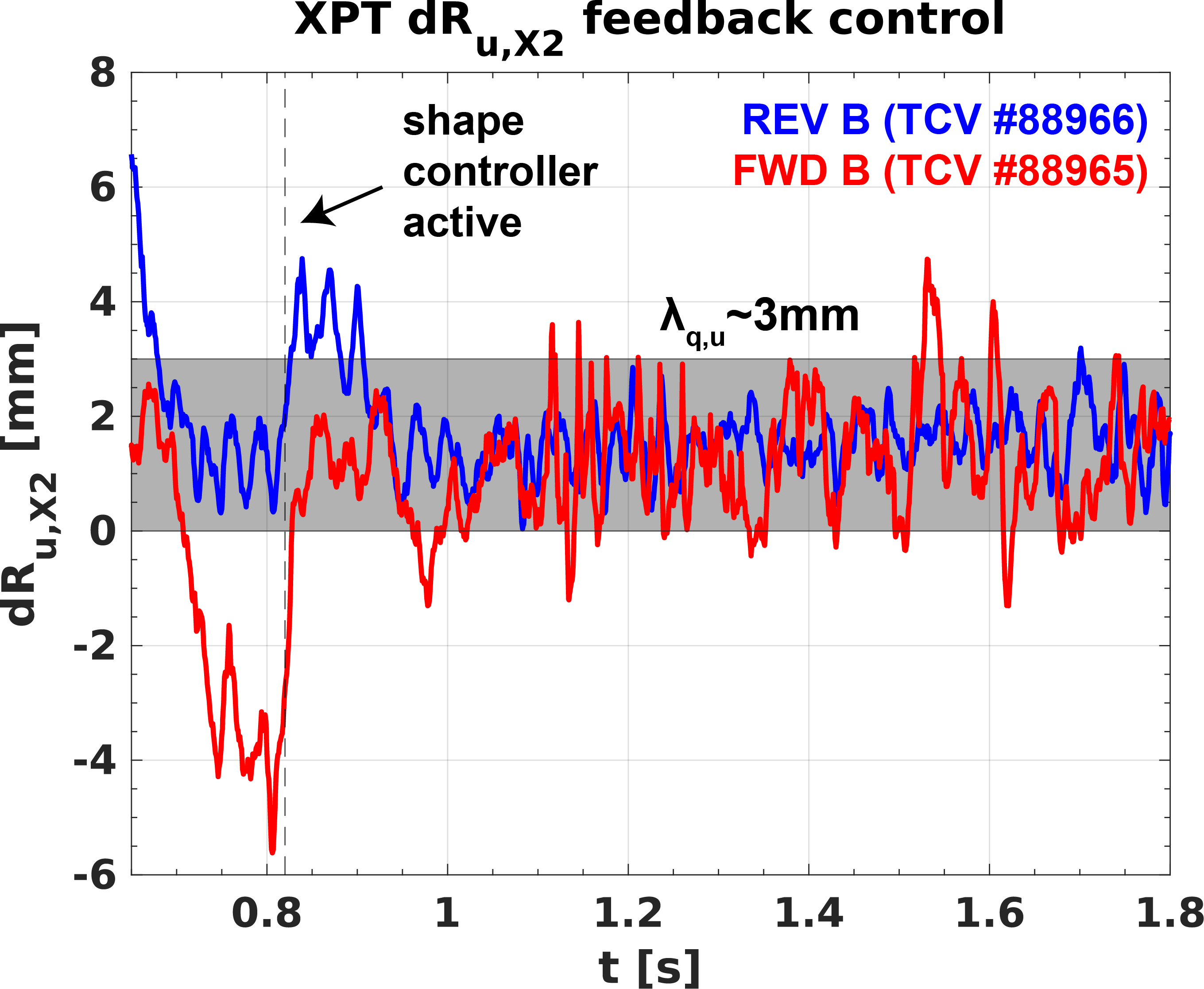}
    \caption{Demonstration of X-point separation control in XPT. Time traces of $\text{d}R_{\text{u},X2}$ obtained from LIUQE reconstruction in two XPT discharges, \#88966 (reversed field) and \#88965 (forward field). The PID shape controller is activated at $t=0.82\,\text{s}$. During the controlled phase, \#88966 achieves $\text{d}R_{\text{u},X2} = 1.53 \pm 0.15 \, \text{mm}$, whereas \#88965 exhibits a more fluctuating $\text{d}R_{\text{u},X2}=1.27 \pm 1.01 \,\text{mm}$.}
    \label{fig:shape_ctrl}
\end{figure}

Most previous TCV experiments relied on precomputed poloidal field coil current traces \cite{hofmann_fbt_1988} to generate an XPT equilibrium, requiring iterations to obtain the desired X-point separation. In the experiments described here, the X-point separation is feedback-controlled using a recently developed plasma shape controller \cite{mele_design_2025}, which enables non-iterative matching of $\text{d}R_{\text{u},X2}$ in XPT discharges. Time traces of the reconstructed $\text{d}R_{\text{u},X2}$ from two XPT discharges, \#88966 (reversed field) and \#88965 (forward field), are shown in figure \ref{fig:shape_ctrl}. A non-zero flux difference reference is imposed between the two X-points to keep the secondary X-point in the common flux region. Around $200\,\text{ms}$ after the shape controller is activated, $\text{d}R_{\text{u},X2}$ reaches the control target and is sustained within one heat flux decay width ($\lambda_{q,\text{u}}\approx3\,\text{mm}$) throughout the discharge.

\subsection{Boundary and divertor diagnostics}
TCV is equipped with an extensive set of boundary and divertor diagnostics \cite{reimerdes_tcv_2017-1,fasoli_tcv_2020}, allowing detailed characterization of power exhaust in the XPT divertor configuration. High-resolution profiles of the target perpendicular heat flux $q_{\perp,\text{t}}$ at all outer SPs are measured simultaneously at $280\,\text{Hz}$ by a set of three infrared (IR) cameras \cite{zurita_infrared_2026}, shown in figure \ref{fig:eqm}(a) and (b). The horizontal IR system can be moved to the midplane port to measure the inner strike point in dedicated repeat discharges. To ensure good IR measurements, the SPs are positioned to avoid tile gaps. The target parallel heat flux is obtained from $q_{\parallel,\text{t}}=(q_{\perp,\text{t}}-q_{\text{bg}})/\sin{\alpha}$, where $q_{\text{bg}}$ is the background heat flux, assumed to be due to radiation, and $\alpha$ is the grazing angle of the magnetic field line at the tile surface. The effect of $q_{\parallel}$ compression or dilution due to variations in the target radial position $R_{\text{t}}$ can be removed by projecting $q_{\parallel,\text{t}}$ upstream \cite{maurizio_divertor_2018} via
\begin{equation}\label{eq:mapped}
    q_{\parallel,\text{u}}^{*} = q_{\parallel,\text{t}} \cdot \bigg(\frac{R_{\text{t}}}{R_{\text{u}}}\bigg),
\end{equation}
where the ratio is simply the total flux expansion $f_{R,\text{t}}\approx R_{\text{t}}/R_{\text{u}}$, corresponding to the change in flux tube cross-section between the outboard midplane and the target (see section \ref{subsection:Rt_effect}). The multispectral imaging system MANTIS \cite{perek_mantis_2019} captures spectrally filtered images of the divertor region at $200\,\text{Hz}$, which are tomographically inverted to yield two-dimensional (2D) emissivity maps of line emission in the poloidal plane. The CIII ($\text{C}^{2+}$, $465\,\text{nm}$) impurity line emission front \cite{theiler_results_2017,harrison_detachment_2017}, defined by a 50\% drop in emissivity along the divertor leg, serves as a proxy for the low-temperature region ($T_{e}\sim7\,\text{eV}$ \cite{martinelli_spectroscopic_2023}), allowing access to detachment-relevant conditions to be monitored. Finally, the Thomson scattering (TS) diagnostic \cite{blanchard_thomson_2019}, operating at $60\,\text{Hz}$, provides local measurements of electron density and temperature in the divertor region.

\section{Properties of an attached XPT divertor}\label{section:attached}
This section presents the features of the reversed-field XPT divertor during the full-power, unseeded phase, in which the outer strike points are attached. The main observation is the emergence of an unconventional line emission and density pattern, attributed to strong cross-field convection in the divertor. The corresponding heat fluxes at all outer strike points, and their sensitivities to the X-point separation $\text{d}R_{\text{u},X2}$ within $\sim 1\lambda_{q,\text{u}}$, are characterized in detail.
\begin{figure}
    \centering
    \includegraphics[width=1\linewidth]{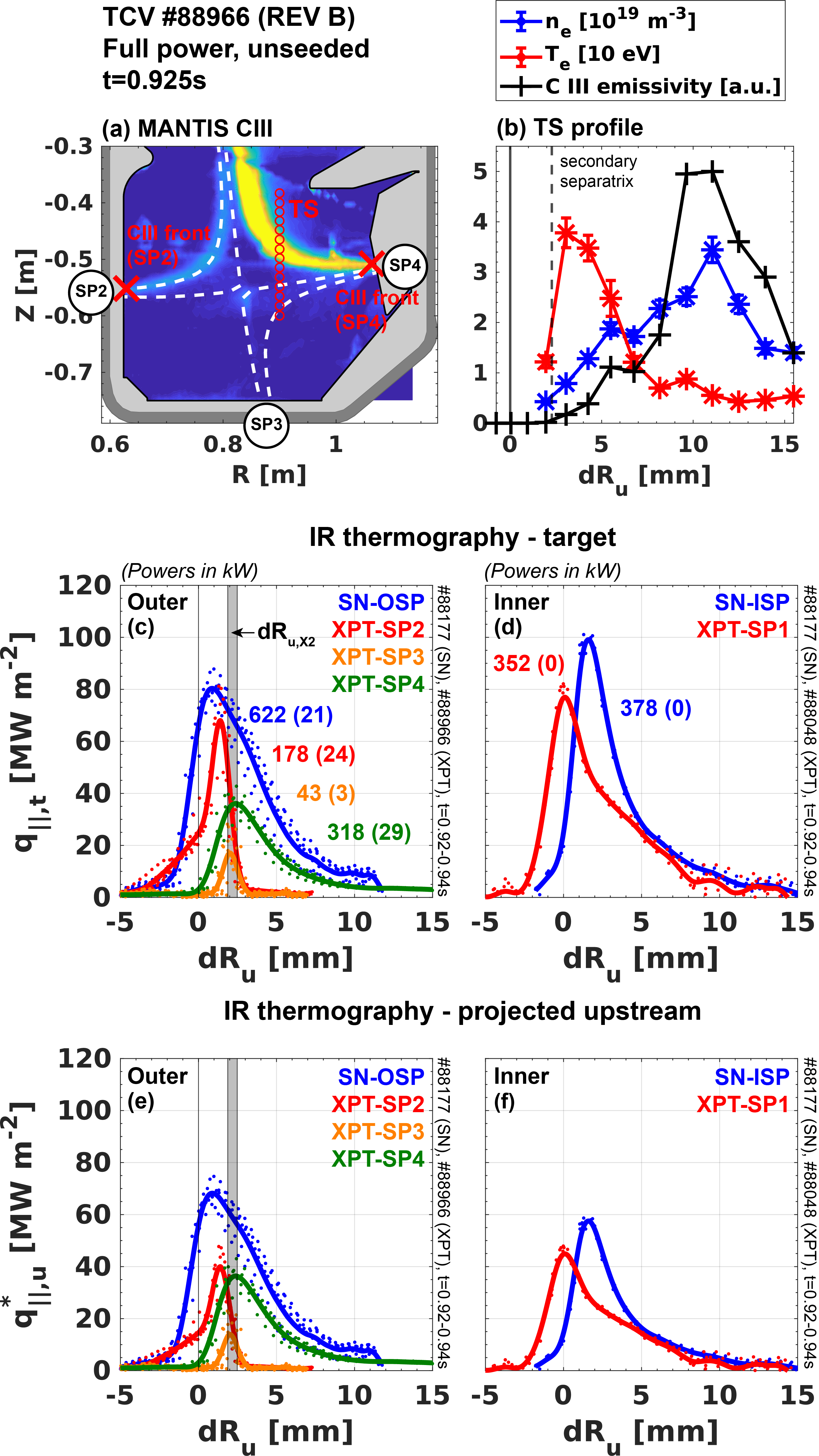}
    \caption{Properties of an attached XPT divertor. (a) CIII emissivity map in the divertor and (b) TS profiles of electron density and temperature, together with the CIII emissivity, at the measurement points marked in (a). (c)--(f) Parallel heat flux profiles at all SN and XPT strike points measured by IR thermography, showing target values at the (c) outer and (d) inner strike points, and upstream-projected values (see equation (\ref{eq:mapped})) at the (e) outer and (f) inner strike points. The integrated power deposited on each strike point (with the background radiation contribution in brackets) is also given.}
    \label{fig:attached}
\end{figure}

\begin{figure*}
    \centering
    \includegraphics[width=0.75\linewidth]{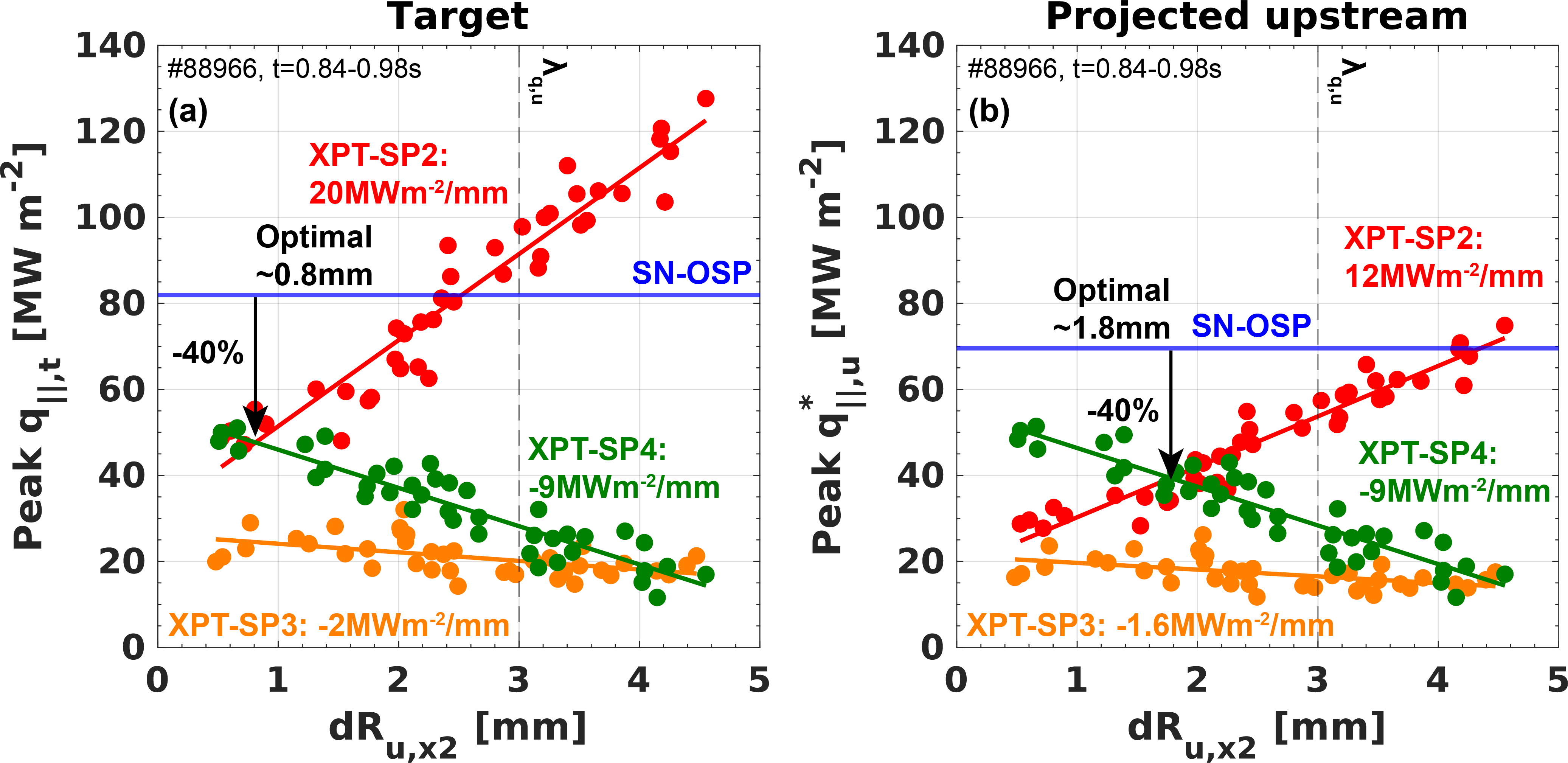}
    \caption{Variation of the peak (a) target and (b) upstream-projected parallel heat flux at the XPT outer strike points (SP2--4) with X-point separation during the unseeded phase in reversed field. Representative SN-OSP values and the estimated sensitivities are also shown.}
    \label{fig:attached_dRx2_scaling}
\end{figure*}
\subsection{Unconventional CIII emission and density distribution}\label{subsection:xpt-structure}

The CIII line emission in the XPT divertor (figure \ref{fig:attached}(a)) shows a ``two-branch'' structure, unlike typical observations in SN discharges, with CIII emission split along the leg and extending to both SP2 and SP4, indicating attached conditions with target electron temperatures high enough to excite $\text{C}^{2+}$ ions. The SP4 branch is a factor of 3 more emissive than the SP2 branch. The two-branch structure also implies two distinct CIII emission fronts, which requires an analysis workflow that tracks each front position separately (\ref{appendix:multi-front}), as marked in figure \ref{fig:attached}(a). Between the SP2 and SP4 emission branches lies a ``hollow'' secondary X-point volume with little emission, apart from some weak emission right at the secondary X-point that does not extend to SP3.

The cause of this unconventional emission pattern appears to be a density redistribution in the XPT divertor, attributed to strong cross-field convective transport. Local TS measurements (figure \ref{fig:attached}(b)) show that the highly emissive SP4 branch coincides with the peak of the divertor electron density profile ($\sim3.5 \times10^{19}\,\text{m}^{-3} \approx 3.5 \cdot n_{e,\text{u}}$). This peak is shifted radially outward from the primary separatrix by almost $4\lambda_{q,\text{u}}$, to $\text{d}R_{\text{u}}\approx11\,\text{mm}$. Such localized peaking suggests that it originates from cross-field convection rather than from diffusive spreading. The region closer to the secondary X-point, at $2.3\, \text{mm}<\text{d}R_{\text{u}}<7\,\text{mm}$, is by contrast hot ($T_{e}\sim40 \,\text{eV}$) and at low density ($n_{e}<2 \times 10^{19} \,\text{m}^{-3}$) with weak emission. A steep electron temperature gradient, from $40\,\text{eV}$ to $10\,\text{eV}$, is also observed across the secondary separatrix. TS indicates that the plasma density below the secondary separatrix is below the detection threshold. Although TS measurements are not available at the SP2 branch, its weak emission may be caused by relatively low electron and $\text{C}^{2+}$ ion densities and higher temperatures than in the emissive part of SP4.

\subsection{Heat flux profiles at the XPT strike points}

The measured target parallel heat flux profiles $q_{\parallel,\text{t}}$ are shown in figure \ref{fig:attached}(c). Despite its weak CIII emission, SP2 exhibits high target heat fluxes, comparable to those at SN-OSP. Note that this occurs at $\text{d}R_{\text{u},X2}\approx 2.2\pm0.3\,\text{mm}$, well within one $\lambda_{q,\text{u}}$, implying that only a small fraction of the SOL is magnetically connected to SP2 and that the connection length is maximized within the power-carrying channel. The high heat flux at SP2 is partly attributed to its small $R_{\text{t}}$, which increases the heat flux by $67\%$ with respect to the upstream value (equation (\ref{eq:Rt_effect})). The main cause, however, is argued in sections \ref{subsection:asym} and \ref{subsection:drx2_breakdown} to be the impeded power dissipation in the SP2 branch. Removing the geometrical effect of total flux expansion using equation (\ref{eq:mapped}), i.e.\ expressing the heat flux as $q_{\parallel,\text{u}}^{*}$, reveals a net $40\%$ reduction in peak parallel heat flux across both SP2 and SP4 compared with the SN (figure \ref{fig:attached}(e)). The disconnected SP3, despite showing neither detectable CIII emission nor a TS signal, is found to be activated in the IR measurements, exhibiting a clear hot spot with a $q_{\parallel,\text{u}}^{*}$ profile that peaks at a value a factor of 2 lower than at SP2 and SP4.

%The power deposited on the strike point is computed by spatially integrating $q_{\perp,\text{t}}$ (including the radiation contribution) over the range $-5\,\text{mm}\leq\text{d}R_{\text{u}}\leq15\,\text{mm}$, which is estimated to be $620\,\text{kW}$ at SN-OSP. Although XPT-SP2 yields peak parallel heat fluxes comparable to the SN, its total power is much smaller ($180\,\text{kW}$), a result of the smaller strike point radius and the narrow hot spot formed by only a small fraction of the SOL. 

Increasing the connection length of the outer leg may redirect power to the inner strike point, as shown in \cite{maurizio_conduction-based_2019,maurizio_effect_2019}. In the present experiment, no heat flux penalty is observed at the inner strike point (XPT-SP1) during the pre-detachment phase when the outer divertor connection length is doubled. In fact, XPT-SP1 (figure \ref{fig:attached}(d) and (f)) exhibits a 20\% lower peak $q_{\parallel,\text{t}}$ than SN-ISP, and its heat flux profile appears shifted radially inward. The exact mechanisms behind the in-out power sharing in the XPT may depend on the SOL regime and on the character of ballooning transport at the outboard midplane \cite{corsaro_impact_2025,carpita_investigation_2026}.
% No correlation between $\text{d}R_{\text{u},X2}$ and XPT-SP1 peak $q_{\parallel,\text{t}}$ is observed. 

\subsection{Sensitivity of XPT peak target heat fluxes to X-point separation}\label{subsection:attached_dRx2}

The relationship between the peak target parallel heat flux (both $q_{\parallel,\text{t}}$ and $q_{\parallel,\text{u}}^{*}$) at all XPT outer strike points and the X-point separation $\text{d}R_{\text{u},X2}$ is shown in figure \ref{fig:attached_dRx2_scaling}. This analysis is based on LIUQE reconstructions that have previously been shown to resolve the flux difference between X-points accurately enough to reveal physical dependencies that were independently confirmed by other diagnostics \cite{maurizio_effect_2019,fevrier_detachment_2021,lafferty_fast_2026}, though systematic uncertainties in $\text{d}R_{\text{u},X2}$ may be present \cite{carpita_physics_2026,carpita_investigation_2026,carpita_inprep}. During the unseeded phase ($t=0.84\text{--}0.98\,\text{s}$), uncontrolled fluctuations of $\text{d}R_{\text{u},X2}$ between 0 and $5\,\text{mm}$, and the corresponding variations in peak parallel heat flux, allow this sensitivity to be quantified. The dependence of peak $q_{\parallel,\text{t}}$ ($q_{\parallel,\text{u}}^{*}$) on $\text{d}R_{\text{u},X2}$ within $\sim 1 \lambda_{q,\text{u}}$ is well described by linear regression, and the sensitivity in absolute terms is quantified by the slope of the fitted line.

In terms of target values, a reduction of the peak $q_{\parallel,\text{t}}$ below the SN level ($\approx 80\,\text{MW m}^{-2}$) occurs only for $\text{d}R_{\text{u},X2}< 2.5\,\text{mm}$ ($\sim0.8\lambda_{q,\text{u}}$), with an optimum at $\text{d}R_{\text{u},X2}\approx0.8\,\text{mm}$  ($\sim0.3\lambda_{q,\text{u}}$), indicating a rather narrow range of $\text{d}R_{\text{u},X2}$ where the XPT performs better than the SN (figure \ref{fig:attached_dRx2_scaling}(a)). This range is wider when the values are projected upstream. A reduction of $q_{\parallel,\text{u}}^{*}$ below the SN level is achieved for $\text{d}R_{\text{u},X2}< 4\,\text{mm}$ ($\sim 1.25\lambda_{q,\text{u}}$), with an optimum at $\text{d}R_{\text{u},X2}\approx1.8\,\text{mm}$ ($\sim 0.6\lambda_{q,\text{u}}$). 

The peak heat flux sensitivity to $\text{d}R_{\text{u},X2}$ at SP2 is found to be a factor of 2.2 higher than at SP4. Removing the $q_{\parallel,\text{t}}\propto 1/R_{\text{t}}$ effect and expressing the results in terms of $q_{\parallel,\text{u}}^{*}$ shows that the sensitivity at SP2 is 33\% higher than at SP4. Taking the heat flux at the optimal point, $\text{d}R_{\text{u},X2}\approx1.8\,\text{mm}$ ($\sim 0.6\lambda_{q,\text{u}}$), as the reference value, the sensitivity can also be expressed in relative terms, as the percentage change in $q_{\parallel,\text{u}}^{*}$ per $\lambda_{q,\text{u}}$, giving an estimated 89\% for SP2 and $-69\%$ for SP4. A higher sensitivity at SP2 has also been predicted by SOLPS-ITER modelling in reversed field \cite{carpita_physics_2026,carpita_investigation_2026,carpita_inprep}.

The SP3 peak heat flux remains largely insensitive to the X-point separation, but its trend has the same sign as at SP4, which hints that the SP3 heat flux may originate from cross-field transport into the SP4-SP3 manifold from SOL flux tubes connected to SP4.
\begin{figure*}
    \centering
    \includegraphics[width=0.6\linewidth]{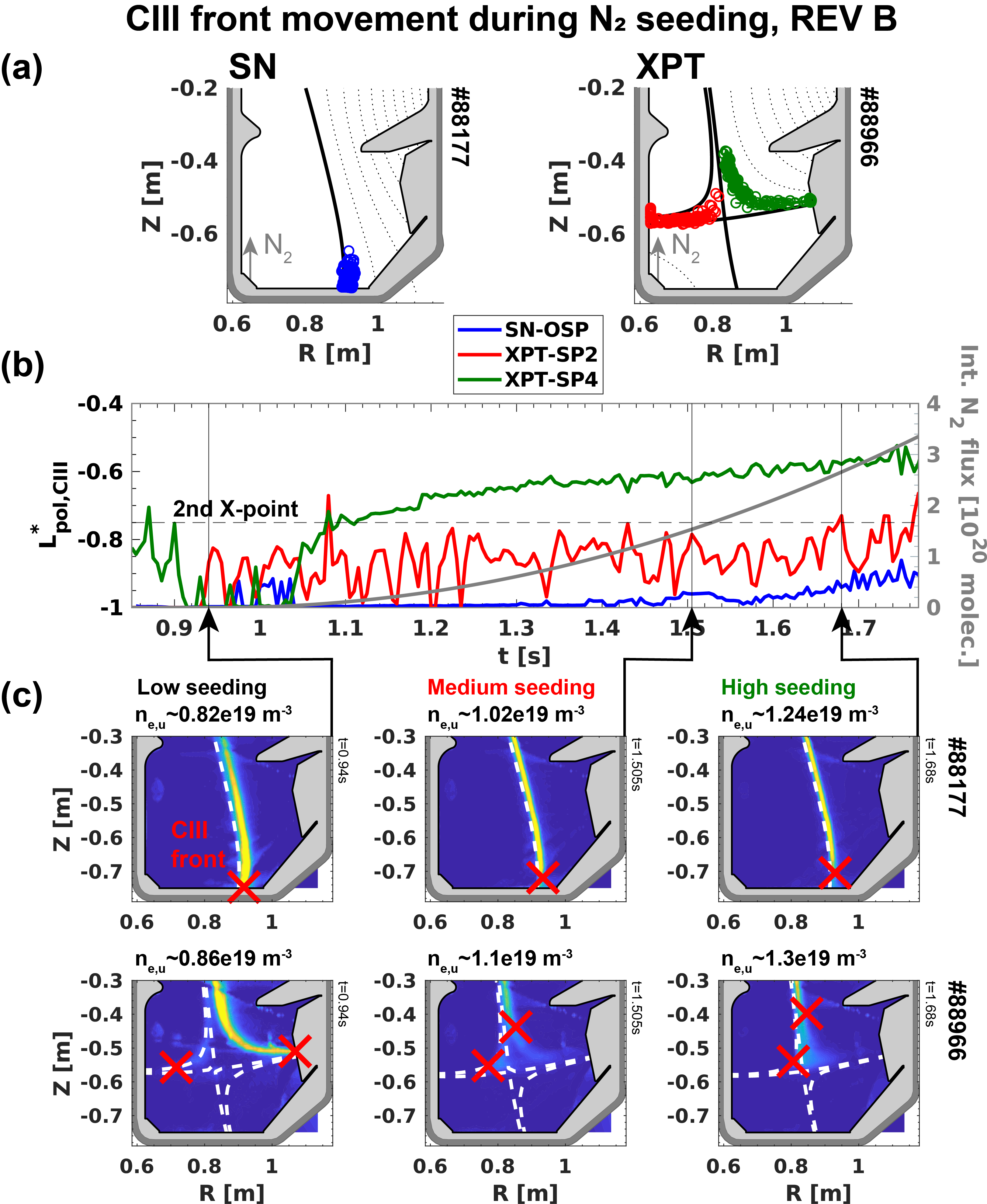}
    \caption{Evolution of the CIII front during nitrogen seeding in the SN and XPT. (a) Position of the CIII front marked in the poloidal plane. (b) Time traces of the normalized CIII front position. (c) CIII emissivity maps at three representative time points. Based on the time-integrated nitrogen flux, three phases are defined as low seeding ($0\text{--}1.1\times10^{20} \,\text{molec.}$), medium seeding ($1.1\text{--}2.2\times10^{20} \,\text{molec.}$), and high seeding ($2.2\text{--}3.3\times10^{20} \,\text{molec.}$).}
    \label{fig:CIII_traces}
\end{figure*}
\section{Detachment of high-power XPT discharges}\label{section:detachment}

In the SN, the high-power scenario has been shown to limit detachment access even with substantial nitrogen seeding, up to the point of unacceptable core contamination \cite{lee_high-power_2026}. This section compares the divertor plasma response to nitrogen seeding in XPT and SN discharges. The key finding is that detachment of the outer divertor is accessed in the XPT, indicating a large reduction of the detachment threshold compared with the SN and thus demonstrating improved power exhaust performance with the XPT geometry. Detailed analysis reveals that detachment evolves asymmetrically across the XPT strike points, with SP4 detaching before SP2. This is related to the unconventional density distribution in the XPT divertor, which drives asymmetric power dissipation by impurity radiation across the outer SOL. The sensitivity of the target conditions (both impurity front position and target heat flux) to the X-point separation is characterized at each strike point and found to decrease at high impurity seeding.

\subsection{Reduction of the detachment threshold in the XPT}\label{subsection:detachment_multi}
The transition to detachment-relevant temperatures in the outer divertor is inferred from the movement of the CIII emission fronts ($T_{e}\sim7\,\text{eV}$). The observations are summarized in figure \ref{fig:CIII_traces}. The poloidal position of the CIII front, normalized to the leg length, is denoted by $L_{\text{pol,CIII}}^{*}$. It ranges from $L_{\text{pol,CIII}}^{*}=-1$ at the target to $L_{\text{pol,CIII}}^{*}=0$ at the primary X-point. In the XPT, a multi-front analysis is carried out to track the evolution of both the SP2 and SP4 fronts, which are then compared with the SN case. No trackable CIII emission is found at SP3. The validity of multi-front tracking is discussed in detail in \ref{appendix:multi-front}. As reported in \cite{lee_high-power_2026}, the CIII front at SN-OSP remains at the target and shows little response to nitrogen seeding (reaching only $L_{\text{pol,CIII}}^{*}\approx -0.9$, figure \ref{fig:CIII_traces}(b)), indicating the lack of transition to detachment. 

The XPT, in contrast, displays substantial changes in the CIII emission pattern (figure \ref{fig:CIII_traces}(c)) and clear movement of the emission fronts (figure \ref{fig:CIII_traces}(a) and (b)), indicating that the divertor targets approach detachment-relevant conditions. The SP4 front moves away from the target immediately upon nitrogen seeding and, from $t=1.1\,\text{s}$, stabilizes at $L_{\text{pol,CIII}}^{*}\approx-0.7$, slightly above the secondary X-point. Further increases in nitrogen seeding drive only slow front movement, up to $L_{\text{pol,CIII}}^{*}\approx-0.55$ by the end of the discharge, resembling the reduced front position sensitivity reported in Ohmic L-mode discharges \cite{lee_x_2025}. The SP2 front shows a net displacement from the target during nitrogen seeding but fluctuates below the secondary X-point (shown in section \ref{subsection:drx2_detach} to be caused by variations in $\text{d}R_{\text{u},X2}$). As nitrogen seeding increases, the radial displacement of the CIII emission in the part of the SOL directed to SP4 (section \ref{subsection:xpt-structure}) becomes progressively less pronounced as the secondary X-point volume gradually becomes more emissive, suggesting that the cross-field convective transport may have weakened. The inner target (SN-ISP and XPT-SP1), by contrast, remains attached throughout the nitrogen seeding phase in both configurations. Overall, these observations show that the impurity seeding threshold for detachment is significantly lower in the XPT than in the SN, constituting a demonstration of improved power handling performance.

\subsection{Asymmetric detachment evolution in the XPT}\label{subsection:asym}
\begin{figure}
    \centering
    \includegraphics[width=1\linewidth]{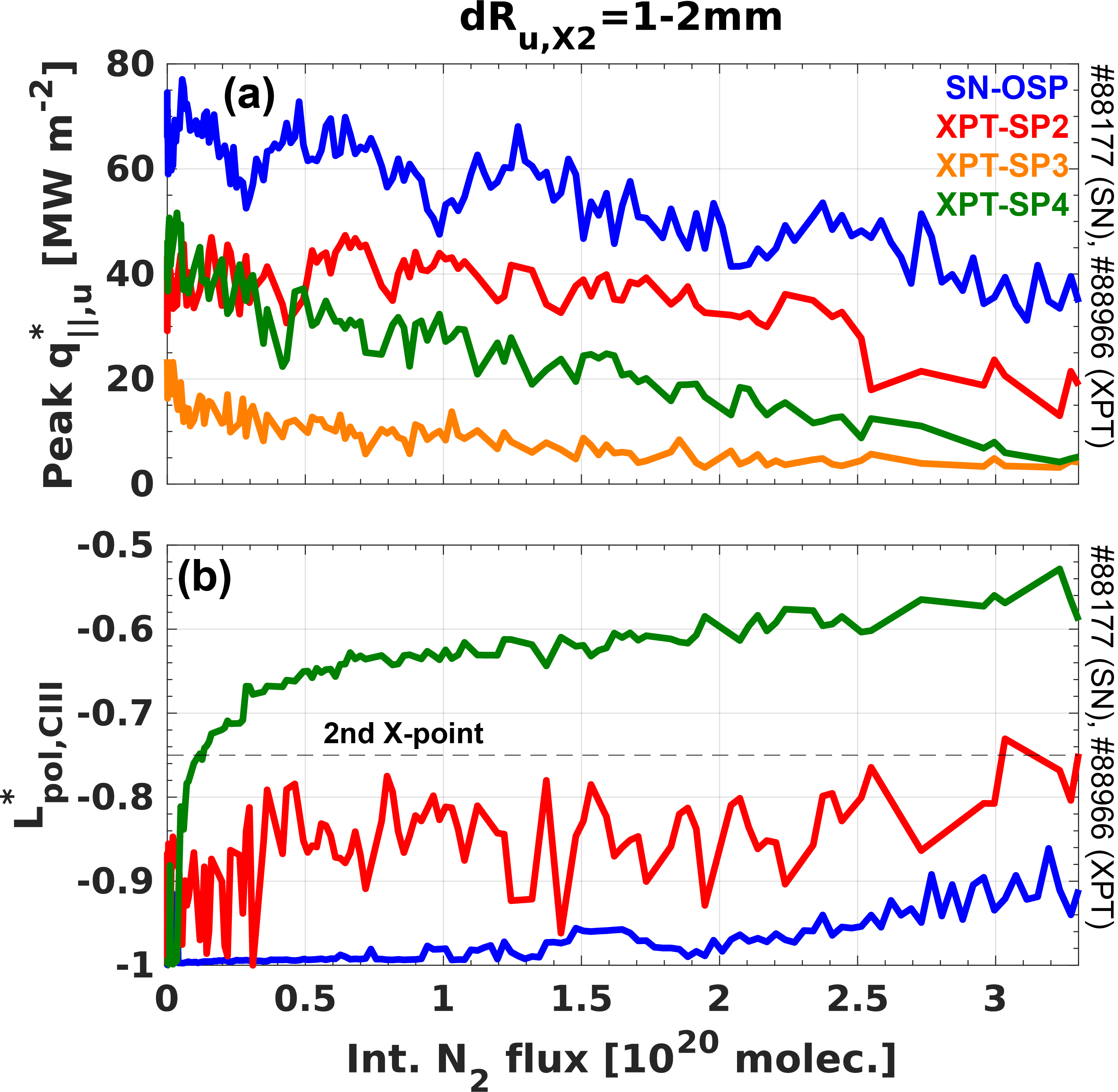}
    \caption{Asymmetric detachment evolution between SP2 and SP4 in the XPT, compared with the SN, showing (a) the peak upstream-projected parallel heat flux and (b) the normalized CIII front position as functions of the time-integrated nitrogen flux at SN-OSP and the XPT outer strike points. Data points are sampled at X-point separations $\text{d}R_{\text{u},X2}=1\text{--}2\,\text{mm}$.}
    \label{fig:evolution_drx2}
\end{figure}

The CIII front traces in figure \ref{fig:CIII_traces}(b) hint that SP4 detaches before SP2. This feature is now investigated in detail. It is shown that both the CIII fronts and the target heat fluxes respond asymmetrically to nitrogen seeding at the XPT outer strike points, with SP4 being much more responsive than SP2. Figures \ref{fig:evolution_drx2}(a) and (b) show peak $q_{\parallel,\text{u}}^{*}$ and $L_{\text{pol,CIII}}^{*}$, respectively, as functions of the time-integrated nitrogen flux at SN-OSP and at all XPT outer strike points. The data points are sampled at $\text{d}R_{\text{u},X2}=1\text{--}2\,\text{mm}$ to exclude the effect of X-point separation variations, which are discussed in section \ref{subsection:drx2_detach}. The peak upstream-projected parallel heat flux $q_{\parallel,\text{u}}^{*}$ at all XPT outer strike points is 40--60\% lower than at SN-OSP throughout the entire nitrogen seeding phase, but each strike point responds differently to seeding. Seeding leads to a $60\%$ drop in peak $q_{\parallel,\text{u}}^{*}$ at SN-OSP, whereas a $90\%$ drop is achieved at SP3 and SP4. Although SP2 receives a lower $q_{\parallel,\text{u}}^{*}$ than SN-OSP, only a 50\% drop is obtained at the highest seeding level, indicating a weaker response to nitrogen seeding than at SP3 and SP4. Moreover, the optimal value $\text{d}R_{\text{u},X2}\approx 1.8\,\text{mm}$ found in figure \ref{fig:attached_dRx2_scaling} no longer holds under nitrogen seeding, since the SP4 peak heat fluxes are consistently lower than those at SP2 (across the entire scanned $\text{d}R_{\text{u},X2}$ range, figure \ref{fig:drx2_detachment}(c) and (d)).

The preferential response to seeding is also seen in the evolution of CIII fronts. The SP4 front moves immediately upon seeding, whereas the movement of the SP2 front is more limited. This asymmetry has an important consequence for the analysis of reversed-field XPT discharges. Detachment at SP4 does not imply detachment of the entire outer divertor, since SP2 can remain attached and continue to receive a significant heat flux.

The asymmetric detachment evolution between SP2 and SP4 occurs at controlled $\text{d}R_{\text{u},X2}\lesssim \lambda_{q,\text{u}}$, where the connection length of the power-carrying flux tubes is similar for both strike points. This suggests that the physics of ``optimized power dissipation'' goes beyond connection length effects (section \ref{subsection:cl_peaking}). A large asymmetry in impurity radiation losses between SP2 and SP4 appears to be the cause, which can be explained using the CIII emission and density profiles observed in figure \ref{fig:attached}, since the impurity radiated power depends quadratically on the local electron density. The asymmetric response to nitrogen seeding is therefore attributed to the high density concentrated in the SP4 branch as a result of cross-field convection, which makes impurity power dissipation there more efficient, whereas radiation at SP2 is impeded by the lower density.

\subsection{Sensitivity of target conditions to X-point separation during nitrogen seeding}\label{subsection:drx2_detach}

\begin{figure*}
    \centering
    \includegraphics[width=0.5\linewidth]{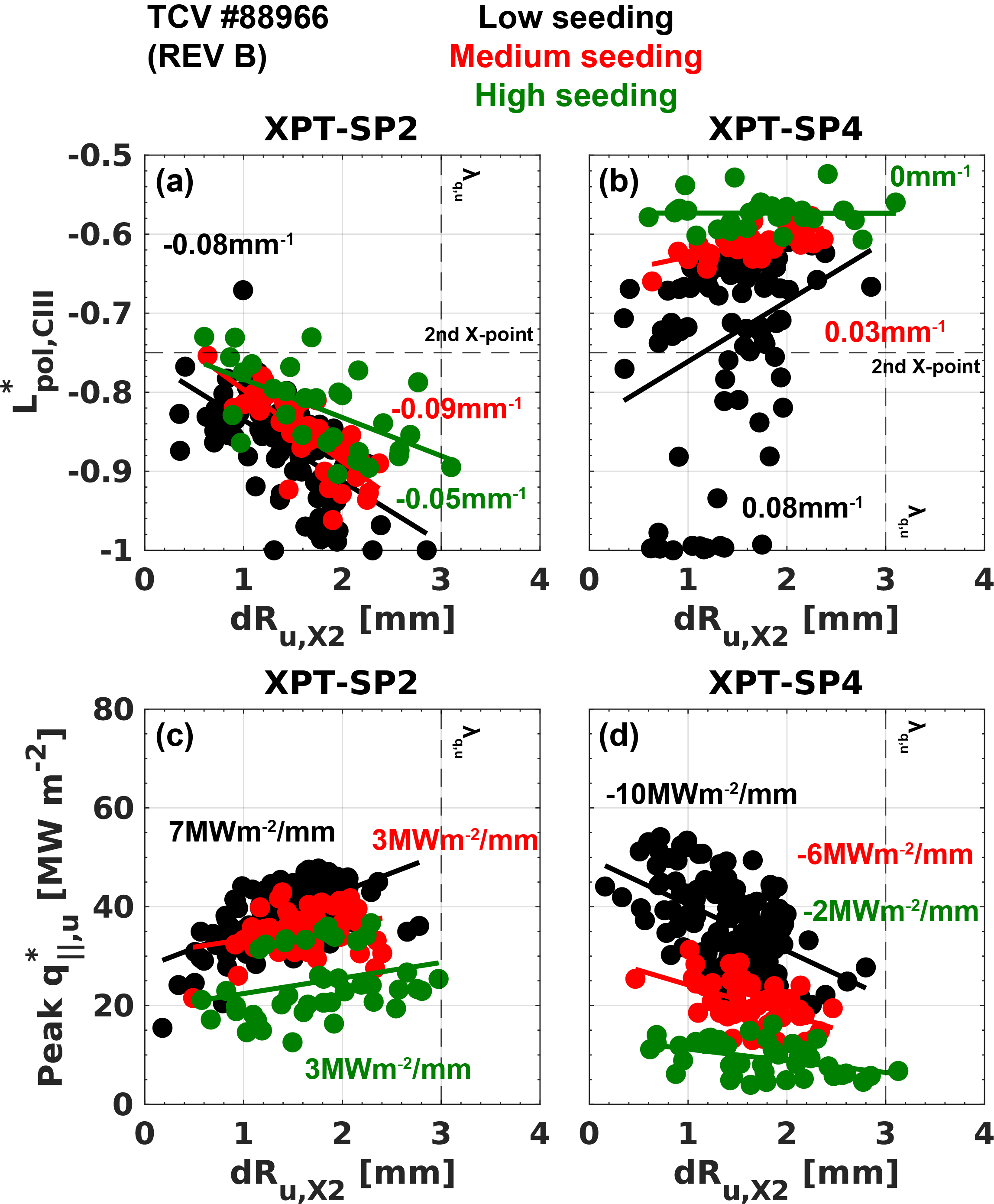}
    \caption{Variation of (a, b) the normalized CIII front position and (c, d) the peak upstream-projected parallel heat flux at the XPT outer strike points, (a, c) SP2 and (b, d) SP4, with X-point separation during the nitrogen seeding phase in reversed field. The three seeding phases are defined as in figure \ref{fig:CIII_traces}. The estimated sensitivities are given in each panel.}
    \label{fig:drx2_detachment}
\end{figure*}

The role of the X-point separation $\text{d}R_{\text{u},X2}$ in the XPT during nitrogen seeding towards the detached regime is investigated here. Both the CIII front position $L_{\text{pol,CIII}}^{*}$ and the peak parallel heat flux $q_{\parallel,\text{u}}^{*}$ vary with $\text{d}R_{\text{u},X2}$. Their dependencies in the three seeding phases are shown in figure \ref{fig:drx2_detachment}. During the shape-controlled phase, $\text{d}R_{\text{u},X2}$ fluctuates between 0 and $3\,\text{mm}$ (i.e.\ within one heat flux decay width $\lambda_{q,\text{u}}$).

The fluctuating SP2 front discussed above is a consequence of $\text{d}R_{\text{u},X2}$ variations (figure \ref{fig:drx2_detachment}(a)), where the tracked front moves away from the target as $\text{d}R_{\text{u},X2}$ decreases. Although the opposite trend is found for the SP4 front (figure \ref{fig:drx2_detachment}(b)), its position quickly becomes insensitive to $\text{d}R_{\text{u},X2}$ once the front moves above the secondary X-point upon nitrogen seeding, so the SP4 CIII front time trace is not oscillatory.

As nitrogen seeding increases, the absolute sensitivity of both peak $q_{\parallel,\text{u}}^{*}$ (target heat flux) and $L_{\text{pol,CIII}}^{*}$ (detachment state) to $\text{d}R_{\text{u},X2}$ decreases. The decrease is generally larger at SP4 than at SP2. For peak $q_{\parallel,\text{u}}^{*}$, the sensitivity at SP2 and SP4 decreases by factors of 2.3 and 5, respectively, from low to high seeding. For $L_{\text{pol,CIII}}^{*}$, the sensitivity at SP2 decreases by a factor of 1.6 at high seeding, while SP4 becomes essentially insensitive to $\text{d}R_{\text{u},X2}$ altogether. The consequences of any uncontrolled displacement of $\text{d}R_{\text{u},X2}$ for XPT power exhaust are therefore correspondingly reduced. The results suggest that the magnetic control requirements on $\text{d}R_{\text{u},X2}$, which could otherwise challenge the practical implementation of the XPT, tend to relax towards strongly seeded, detached regimes.

\section{Field reversal, $E\times B$ drifts, and the origin of the density redistribution in the XPT divertor}\label{section:drifts}
This section addresses the physical origin of the unconventional density redistribution in the XPT divertor, identified in section \ref{section:detachment} as the cause of the asymmetric impurity radiation profile and hence of the asymmetric detachment evolution between the XPT outer strike points. This redistribution was argued in section \ref{subsection:xpt-structure} to be driven by strong cross-field convective transport, which displaces the density peak radially by several heat flux decay widths.

Here, it is shown that this transport originates from $E \times B$ drifts around the secondary X-point, as evidenced by substantial changes in the divertor density profiles in field-reversal experiments. Reversing the $E\times B$ drift direction by changing the magnetic field direction is known to drive in-out asymmetries in plasma density and particle fluxes \cite{fevrier_nitrogen-seeded_2020,colandrea_investigation_2024}. Modifications of the $E \times B$ drift patterns in an XPT divertor were predicted by SOLPS-ITER modelling \cite{carpita_physics_2026,carpita_investigation_2026,carpita_inprep} and are discussed in section \ref{subsection:ExB_modified}. Upon field reversal, the density distribution is found to reverse between SP2 and SP4. The impurity radiation, previously concentrated at SP4 where it favoured heat flux reduction, shifts to SP2 accordingly, swapping the target heat flux profiles between the two strike points.

\subsection{$E\times B$-flux-driven density redistribution in the XPT divertor}
\begin{figure}
    \centering
    \includegraphics[width=0.9\linewidth]{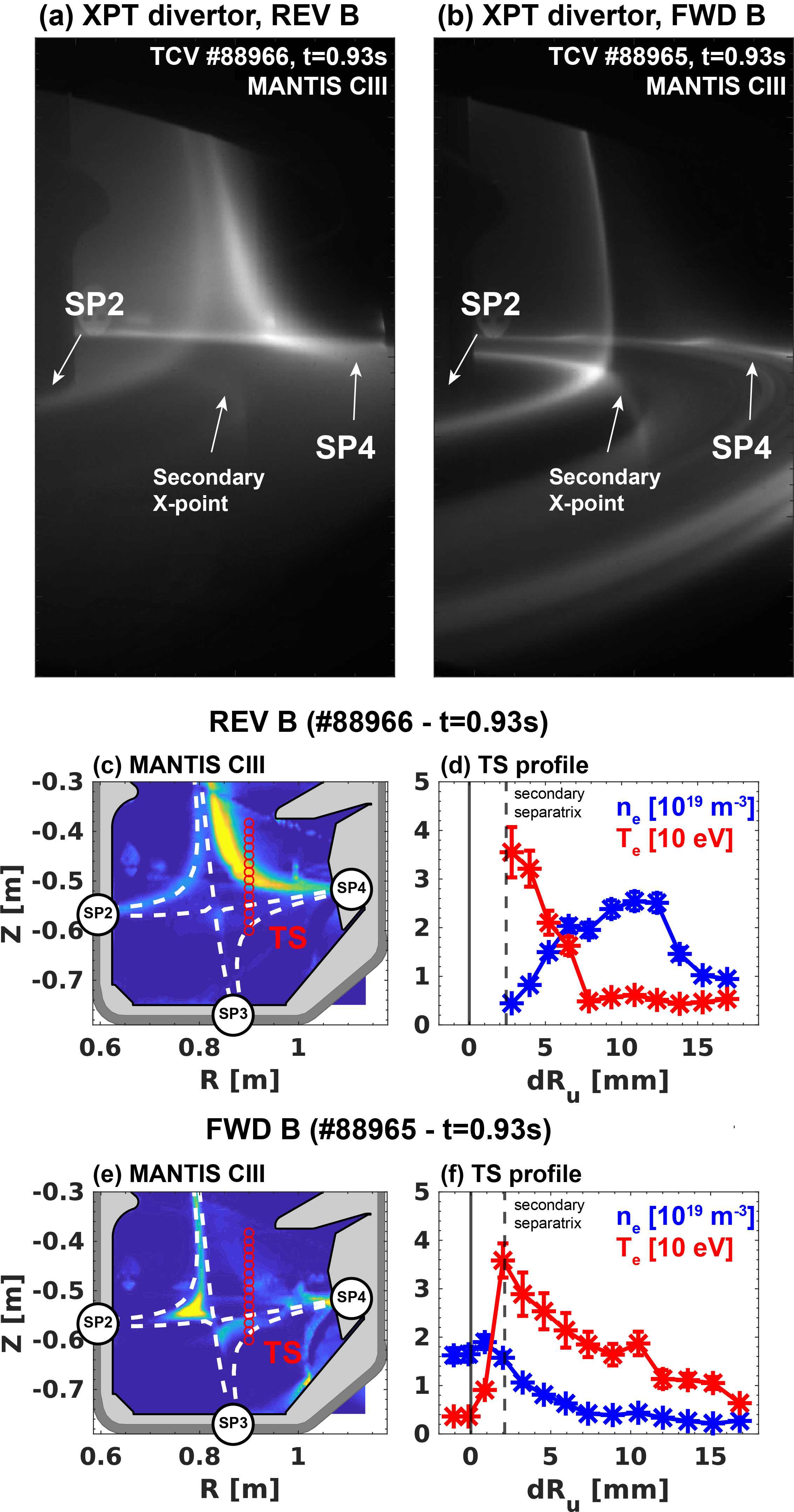}
    \caption{Impact of field reversal on the density distribution in the XPT divertor. CIII-filtered MANTIS images of the lower divertor chamber in (a) reversed and (b) forward field. Reconstructed CIII emissivity maps (c, e) and TS profiles of electron density and temperature (d, f) in reversed (c, d) and forward (e, f) field. Note that the changes are consistent with figure \ref{fig:xpt_physics}(c).}
    \label{fig:CIII_images}
\end{figure}

A high-power L-mode scenario in forward field (TCV \#88965) is obtained and compared with the reversed-field case. The upstream parameters of the scenarios are matched ($I_{\text{P}}=300\,\text{kA}$, $P_{\text{SOL}}\approx 2.5\,\text{MW}$, $n_{e,\text{u}}\approx 0.8 \times10^{19}\,\text{m}^{-3}$). Upon field reversal, changes in the divertor CIII emission are observed (figure \ref{fig:CIII_images}(a) and (b)), indicating a redistribution of density. In reversed field (figure \ref{fig:CIII_images}(a)), the two-branch structure and the hollow secondary X-point are clearly visible, as discussed extensively in section \ref{section:attached}. In the forward-field case (figure \ref{fig:CIII_images}(b)), the secondary X-point volume in the SP2 branch becomes strongly emissive, and the SP2 emission front appears not to extend to the strike point. The bright SP4 branch is no longer present. Instead, weak emission is observed that closely follows the secondary separatrix and extends to SP4. This emission is localized \emph{below} the secondary separatrix, in the secondary PFR (shown more clearly in the reconstructed emissivity map in figure \ref{fig:CIII_images}(e)), and does not connect to the emission extending vertically upward along the divertor leg.

%Field-direction-dependent changes in the divertor line emission on TCV have been previously investigated in Ohmic L-mode discharges, which typically show a radial outward shift of emission in reversed field \cite{colandrea_investigation_2024}. However, such a visually pronounced change with field direction is unprecedented on TCV.

%We focus on the unseeded. The separatrix density $n_{e}=9.25e18$ and $8.5789e18$ and Te = 102.9135. separatrix temperature is somewhat lower $T_{e}=66.6932$

Analysis of the CIII emissivity maps and TS measurements (figure \ref{fig:CIII_images}(c)--(f)) indicates clear changes in the density distribution in the XPT divertor, with the behaviour at SP2 and SP4 essentially swapped. Consistent with the absence of the bright SP4 branch in forward field, the density shift into the far SOL is no longer observed. Instead, the SP4 volume above the secondary separatrix is now at low density ($\lesssim1.5 \times10^{19} \,\text{m}^{-3}$), with a high electron temperature ($\sim20\text{--}40\,\text{eV}$) (figure \ref{fig:CIII_images}(f)). The SP2 manifold features a highly emissive X-point region, likely denser than the rest of the SOL. The emission front is pronounced and poloidally displaced from SP2, suggesting a detached strike point. Furthermore, the reconstructed CIII emissivity map (figure \ref{fig:CIII_images}(e)) clearly shows emission in the upstream-disconnected manifold between SP4 and SP3 in the secondary PFR, where TS detects measurable plasma with a peak density of $n_{e}\approx 2\times10^{19}\,\text{m}^{-3}$ ($\sim 2\cdot n_{e,\text{u}}$) and a low electron temperature ($T_{e} \lesssim 10\,\text{eV}$). These changes closely follow the $E \times B$ drift pattern around the secondary X-point depicted in figure \ref{fig:xpt_physics}(c).

\begin{figure*}
    \centering
    \includegraphics[width=1\linewidth]{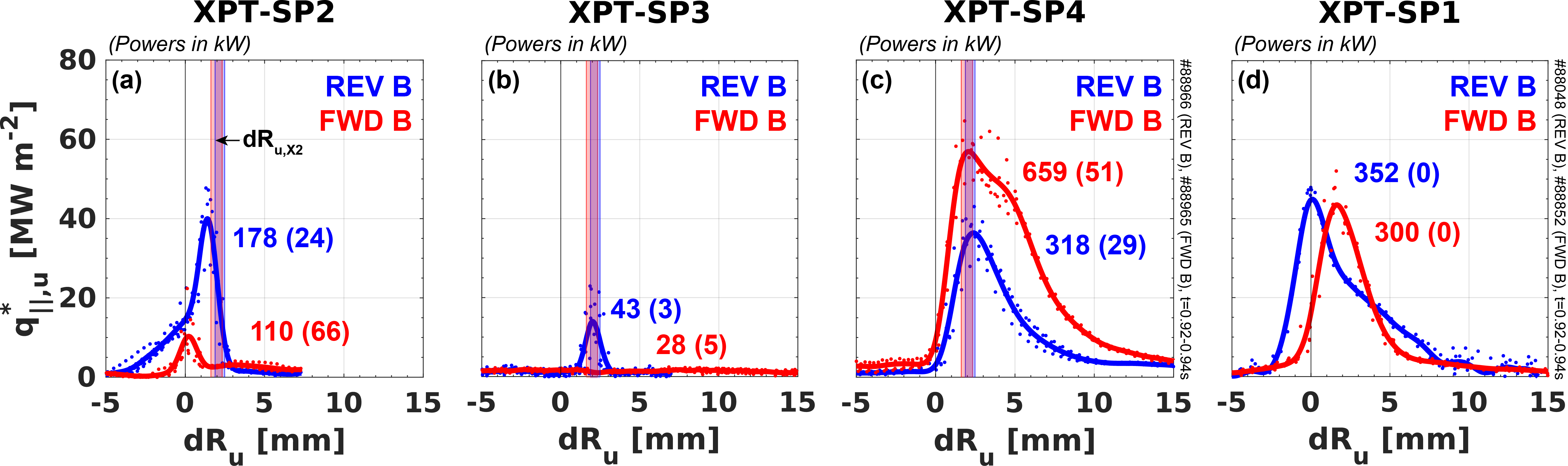}
    \caption{Impact of field reversal on the upstream-projected parallel heat flux profiles at the XPT strike points (a) SP2, (b) SP3, (c) SP4 and (d) SP1. The profiles are taken at matching X-point separations, $\text{d}R_{\text{u},X2}=2.2\pm0.3\,\text{mm}$ (reversed field) and $1.9\pm0.4\,\text{mm}$ (forward field). The integrated
power deposited on each strike point (with the background
radiation contribution in brackets) is also shown.}
    \label{fig:field_rev_heat}
\end{figure*}

\begin{figure}
    \centering
    \includegraphics[width=0.8\linewidth]{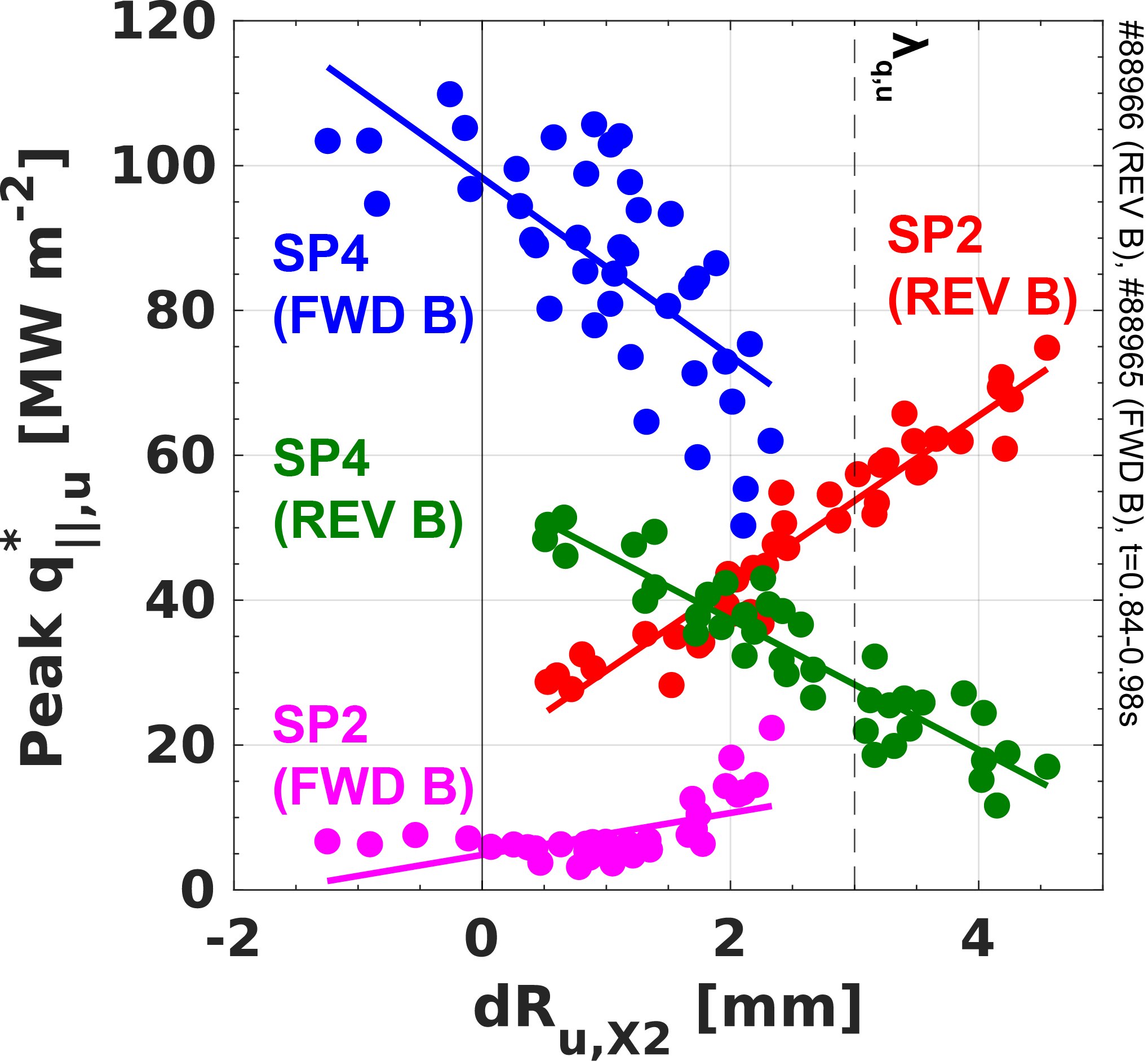}
    \caption{Variation of the peak upstream-projected parallel heat flux at the XPT outer strike points with X-point separation in reversed and forward field during the unseeded phase.}
    \label{fig:field_rev_dRx2}
\end{figure}

\subsection{Consequences of the radiation distribution for power exhaust and X-point separation optimization}\label{subsection:drx2_breakdown}
Since the density redistribution also affects the efficiency of power losses through impurity radiation, the heat flux reaching each strike point is expected to change with the field direction as well. The IR heat flux measurements, summarized in figure \ref{fig:field_rev_heat}, confirm this expectation. Consistent with the observed displacement of the SP2 CIII front, the SP2 peak $q_{\parallel,\text{u}}^{*}$ (figure \ref{fig:field_rev_heat}(a)) drops by 75\% from reversed to forward field. The reduction of plasma density in the SP4 manifold, on the other hand, leads to a 50\% increase in the SP4 peak heat flux (figure \ref{fig:field_rev_heat}(c)). The relative proximity of the two strike points to detachment also reverses with the field direction, with SP2 now detached and SP4 attached. Intriguingly, this reversal also extends to SP3. While a detectable heat flux is found at SP3 in reversed field (figure \ref{fig:field_rev_heat}(b)), this hot spot disappears in forward field. This is interpreted as $E \times B$ convection into the secondary PFR enhancing the dissipation of the SP3 heat flux.

These observations are accompanied by only a small change in the heat flux at SP1. Figure \ref{fig:field_rev_heat}(d) shows the upstream-projected parallel heat flux profile at SP1 for both field directions. The two profiles have similar shapes and peak values, suggesting that the large changes in outer divertor behaviour are dominated by processes specific to the XPT geometry.

The flip in the asymmetry of the XPT target heat flux distribution with field reversal has implications for the role of the X-point separation. Figure \ref{fig:field_rev_dRx2} shows peak $q_{\parallel,\text{u}}^{*}$ at SP2 and SP4 as a function of $\text{d}R_{\text{u},X2}$ for both reversed and forward field. In forward field, data are only available for $-1.5\,\text{mm} \leq \text{d}R_{\text{u},X2} < 2.5\,\text{mm}$. For $\text{d}R_{\text{u},X2}=0\text{--}1.8\,\text{mm}$, the SP2 peak parallel heat flux is reduced to $\leq5\,\text{MW m}^{-2}$, compared with $30\text{--}40\,\text{MW m}^{-2}$ in reversed field, and shows essentially no response to $\text{d}R_{\text{u},X2}$ up to $1.8\,\text{mm}$, beyond which the heat flux starts to increase. This behaviour is attributed to the concentration of plasma density in the SP2 manifold, which enhances its power dissipation efficiency and leaves more room to increase $\text{d}R_{\text{u},X2}$ while sustaining a low heat flux at SP2. SP4 in forward field, on the other hand, has a peak heat flux elevated by almost a factor of 2 with respect to the reversed-field case, and shows stronger sensitivity to $\text{d}R_{\text{u},X2}$. This hints that the asymmetry in X-point separation sensitivities of SP2 and SP4 is reversed in forward field, as also predicted by SOLPS-ITER modelling \cite{carpita_physics_2026,carpita_investigation_2026,carpita_inprep}. An ``optimal'' $\text{d}R_{\text{u},X2}$ for the unseeded forward-field XPT is larger than in the reversed-field case, but the difference cannot be quantified with currently available data.

\section{Discussion and implications for reactors}\label{section:discussion}
This section discusses the implications of the experimental results. The trends found in the present experiments, in particular the role of $E \times B$ drifts, largely confirm the expectations from SOLPS-ITER modelling with drifts \cite{carpita_physics_2026,carpita_investigation_2026,carpita_inprep}. Many of the main conclusions on the reactor extrapolation of the XPT, derived from the modelling work \cite{carpita_physics_2026,carpita_investigation_2026,carpita_inprep}, are therefore reiterated here and supported by the experimental observations presented in this paper.

\subsection{Improved power exhaust performance}
Improvements in divertor power exhaust performance with the XPT configuration over a conventional SN are demonstrated in a high-power TCV scenario, in which the detachment scaling, as measured by the Lengyel metric, lies in the reactor-relevant range \cite{lee_high-power_2026}. Here, the XPT facilitates detachment access, whereas detachment access is limited in the SN. The absolute sensitivity of the divertor conditions to the X-point separation $\text{d}R_{\text{u},X2}$ decreases with impurity seeding (at least within a range of $\sim1\lambda_{q,\text{u}}$), suggesting less stringent magnetic control requirements in the detached XPT regime. While these XPT results are encouraging, the physical mechanism behind the improved performance appears to be complex in the presence of $E\times B$ drift transport inherent to the secondary X-point, and must be carefully understood in order to optimize the XPT and exploit its full power exhaust potential.

\subsection{Role of drift transport}
This work shows experimentally that $E \times B$ drifts significantly change the divertor power exhaust behaviour, and their effects are likely amplified by the lower collisionality and higher temperature of the SOL in the high-power scenario compared with Ohmic L-mode. Distinct drift convection channels can arise from the XPT's magnetic topology and redistribute the density in the XPT divertor, as predicted by SOLPS-ITER modelling \cite{carpita_physics_2026,carpita_investigation_2026,carpita_inprep}. In these experiments, $E \times B$ convection between the SP2 and SP4 branches leads to an asymmetry in impurity radiative dissipation and detachment evolution across the strike points. Such large displacements of particles around the secondary X-point challenge the concept of X-Point Target radiator \cite{lee_x_2025,labombard_adx_2015}, which rely on enhanced radiative losses due to large poloidal flux expansion. Drifts can deplete particles near the secondary X-point (figure \ref{fig:CIII_images}(c)--(f)) in one of the strike point branches and may counteract these geometric effects \cite{carpita_physics_2026,carpita_investigation_2026,carpita_inprep}---with the drifts themselves, in turn, driven by fundamental features of the magnetic geometry.

\subsection{High-field tokamak reactors}
Future high-field tokamaks operating at $B_{0}>10\,\text{T}$ (e.g.\ SPARC, ARC) may exhibit weaker drift effects, since the drift velocity scales as $v_{E\times B} \propto 1/B$. However, the contribution of drift transport relative to other transport mechanisms in the SOL scales approximately as $T_{e}^{3/4}/\sqrt{B \cdot L_{\nabla} }$ \cite{chankin_classical_1997,colandrea_investigation_2024}, where $L_{\nabla}$ is the characteristic gradient scale length in the divertor in the radial (especially across the secondary PFR-SOL boundary, see section \ref{subsection:ExB_modified}) and poloidal directions. Since $T_{e}$ is expected to increase and $L_{\nabla}$ to decrease in a reactor SOL, drift effects at high magnetic field cannot be safely assumed to be negligible, and they can affect plasma and divertor operation.

\subsection{Detachment transition dynamics}
Operating in a deeper detachment state, e.g.\ at higher divertor neutral pressure or with stronger impurity seeding, may suppress drift transport \cite{rizkallah_enhanced_2026,kaveeva_solps-iter_2020,jaervinen_impact_2019}, since the temperature and plasma potential gradients in the divertor flatten. Some signs of reduced drift transport are observed during nitrogen seeding in this work (section \ref{subsection:detachment_multi}). Detached operation will, in any case, be required for steady-state reactor operation. In such detached states, the XPT has demonstrated high resilience to perturbations around the secondary X-point \cite{lee_x_2025,winkel_detachment_2026} and, as shown in this work, to variations in X-point separation. An open question is then the dynamics of the detachment transition in the XPT, which involves the nonlinear interplay between the divertor conditions and $E \times B$ drifts \cite{jaervinen_e_2018,wensing_x-point_2020,yang_dynamical_2026}. High-fidelity simulations have improved the understanding of both stationary and time-dependent drift effects on detachment, especially in X-point geometries \cite{carpita_physics_2026,carpita_investigation_2026,carpita_inprep,yang_dynamical_2026}. Further exploitation of these tools to model drifts in the XPT divertor self-consistently will be important for enabling a realistic assessment for a reactor.

\section{Conclusions}\label{section:conclusion}
The X-Point Target (XPT) is a promising alternative divertor configuration that has shown improved power exhaust performance and is being considered for future tokamak reactors. Reliable extrapolation of the XPT to a reactor requires experiments that demonstrate its effectiveness at SOL parameters approaching reactor values, as well as an understanding of the processes underlying its improved performance. To this end, XPT experiments were conducted in TCV at the highest available ECRH power, with the Lengyel detachment scaling metric reaching reactor-level values. Interpreting the experimental results in the XPT requires simultaneous consideration of the different aspects of the X-point separation and their impact on power exhaust, including SOL splitting, connection length peaking, $E\times B$ drifts around the secondary X-point, and vessel geometry, as summarized in figure \ref{fig:xpt_physics}.

In terms of absolute performance, the XPT in this high-power TCV scenario, operated in reversed field, reduces the peak divertor parallel heat flux below the SN level by 40--60\% throughout the entire discharge, both before and during nitrogen seeding. According to the Lengyel metric, the detachment threshold of the high-power scenario is a factor of 10 higher than in the previously studied Ohmic L-mode and NBI H-mode scenarios, and detachment in the SN is prevented. The most encouraging result is the transition to detachment-relevant conditions in the XPT, characterized by the movement of impurity emission fronts. The sensitivity of the heat flux sharing between the outer strike points to the X-point separation is quantified and found to decrease with increasing impurity seeding, implying that the precision requirements on magnetic control are relaxed in the detached state.

A key observation revealing the processes underlying XPT performance is an unconventional density redistribution around the secondary X-point. A large radial displacement of the divertor plasma density, attributed to cross-field convection, concentrates density in one of the two outer SOL branches, depending on the field direction. As a consequence, impurity radiative dissipation across the outer SOL becomes highly asymmetric, since the power dissipation efficiency scales quadratically with the local electron density. A large asymmetry in detachment evolution therefore emerges upon nitrogen seeding in reversed field, with the high-density branch detaching before the underdense branch.

The origin of the cross-field convection and density redistribution in the XPT is identified as $E \times B$ drift convection inherent to the magnetic topology of the secondary X-point, confirming predictions by SOLPS-ITER modelling \cite{carpita_physics_2026,carpita_investigation_2026,carpita_inprep}. Upon field reversal, the density in the two outer SOL branches flips, together with the radiation distribution, and the direction of density shift is consistent with the expected change in $E \times B$ drift directions. Consequently, the distribution of parallel heat flux between the strike points is also reversed. Due to this directional shift of density and impurity radiation across the SOL, the optimal X-point separation changes accordingly.

The results of this work demonstrate the enhanced power handling capability of the XPT divertor at a reactor-relevant Lengyel metric while also providing insights into its fundamental physics that must be understood for its optimization. The most complex aspect is the appearance of distinct $E \times B$ drift patterns in the XPT geometry, which are shown to be important and are potentially non-negligible at reactor parameters. In the future, these effects should be investigated via dedicated experiments and self-consistent modelling to extrapolate the XPT reliably as a power exhaust solution for tokamak fusion reactors.

\section*{Acknowledgment}
Insightful discussions with S.~Ballinger, M.~Bernert, T.~Body, L.~Corsaro, T.~Eich, A.~McLean, P.~Muscente, J.-S.~Park and M.~Wigram are gratefully acknowledged. This work was supported in part by the Swiss National Science Foundation. This work has been carried out within the framework of the EUROfusion Consortium, partially funded by the European Union via the Euratom Research and Training Programme (Grant Agreement No 101052200 — EUROfusion). The Swiss contribution to this work has been funded in part by the Swiss State Secretariat for Education, Research and Innovation (SERI). Views and opinions expressed are however those of the author(s) only and do not necessarily reflect those of the European Union, the European Commission or SERI. Neither the European Union nor the European Commission nor SERI can be held responsible for them. 

\appendix
\section{Interpretation of multi-front analysis}\label{appendix:multi-front}
CIII front analyses of the XPT have been carried out in previous work \cite{theiler_results_2017,raj_improved_2022,lee_x_2025,winkel_detachment_2026}. In this work, a multi-front analysis is employed to track the multiple CIII fronts in the XPT. The extinction of CIII emission near the target and the movement of the CIII front are generally good indicators of divertor cooling, since the relative abundance of the $\text{C}^{2+}$ charge state depends strongly on temperature. However, because of the non-standard workflow used for multi-front analysis in the XPT divertor, the physical interpretation of the CIII front position is not straightforward.

Front tracking requires defining a region of interest, characterized by a chosen SOL flux surface and a finite leg width. Since most previous CIII front analyses concern a single divertor leg, the region of interest is typically defined to cover the full width of the leg emission, and the front analysis is treated as a 1D problem. In the XPT, with two distinct fronts, the flux surface and leg width must be chosen carefully to track the SP2 and SP4 fronts independently. This treatment ignores the inherent radial structure of the divertor leg emission, meaning that the tracked XPT fronts may not be physically or quantitatively equivalent to the well-studied single-leg CIII front \cite{martinelli_spectroscopic_2023}.

\begin{figure*}
    \centering
    \includegraphics[width=0.8\linewidth]{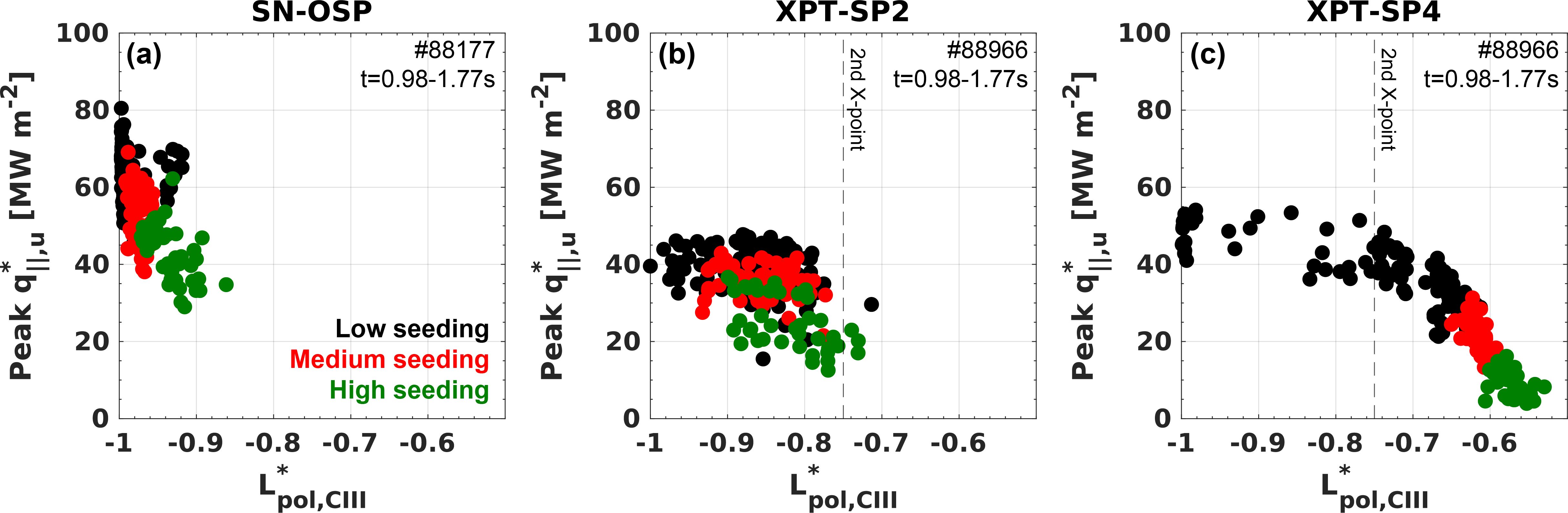}
    \caption{Peak upstream-projected parallel heat flux as a function of the normalized CIII front position in the reversed-field scenario, in three nitrogen seeding phases (defined as in figure \ref{fig:CIII_traces}), measured at (a) SN-OSP, (b) XPT-SP2, and (c) XPT-SP4.}
    \label{fig:ciii_qpar}
\end{figure*}

In this appendix, the significance of the tracked XPT fronts is assessed by correlating the front position with the peak upstream-projected parallel heat flux at each strike point (figure \ref{fig:ciii_qpar}). The observations are summarized in the following list.
\begin{itemize}
    \item At SN-OSP (figure \ref{fig:ciii_qpar}(a)), front movement from $L_{\text{pol,CIII}}^{*}=-1$ to $-0.9$ corresponds to a drop of roughly $60\%$ in peak $q_{\parallel,\text{u}}^{*}$.
    \item At XPT-SP2 (figure \ref{fig:ciii_qpar}(b)), CIII front oscillations within $-1 \leq L_{\text{pol,CIII}}^{*} \leq -0.8$ produce only a small reduction in peak $q_{\parallel,\text{u}}^{*}$, which remains around $40\,\text{MW m}^{-2}$, indicating that this range of $L_{\text{pol,CIII}}^{*}$ does not correspond to detachment onset at SP2. Only at high seeding levels, as the front approaches the secondary X-point, does peak $q_{\parallel,\text{u}}^{*}$ drop by $50\%$.
    \item At XPT-SP4 (figure \ref{fig:ciii_qpar}(c)), peak $q_{\parallel,\text{u}}^{*}$ decreases by up to $88\%$ in the range $-0.7\leq L_{\text{pol,CIII}}^{*} \leq-0.55$, as the front moves slowly above the secondary X-point.
\end{itemize}

It is concluded that, in this XPT dataset, the heat flux reduction occurs when the XPT front position $L_{\text{pol,CIII}}^{*}$ approaches or moves above the secondary X-point. Therefore, although the tracked XPT fronts may not be exactly equivalent to the conventional single-leg front, they remain a useful proxy for proximity to the detached state.

\DeclareFieldFormat[article]{title}{}
\printbibliography

\end{document}